\documentclass[journal]{IEEEtran}

\usepackage{cite}
\usepackage{amsmath,amssymb,amsfonts}
\usepackage{algorithmic}
\usepackage{graphicx}
\usepackage{textcomp}
\usepackage{comment}
\usepackage{multirow}
\usepackage[flushleft]{threeparttable}

\usepackage{siunitx}
\usepackage{enumerate}
\usepackage{booktabs}
\usepackage{subcaption}
\usepackage{hyperref} 
\usepackage[nolist]{acronym}
\usepackage{paralist}
\usepackage{orcidlink}

\usepackage{pifont}
\newcommand{\cmark}{\ding{51}}
\newcommand{\xmark}{\ding{55}}

\begin{document}

\title{PuLsE: Accurate and Robust Ultrasound-based Continuous Heart-Rate Monitoring on a Wrist-Worn IoT Device}

\author{
\IEEEauthorblockN{
    Marco Giordano, \textit{Graduate Student Member, IEEE},
    Christoph Leitner, \textit{Member, IEEE},
    Christian Vogt, \textit{Member, IEEE},
    Luca Benini, \textit{Fellow, IEEE} and 
    Michele Magno, \textit{Senior Member, IEEE}}

\thanks{Marco Giordano, Christian Vogt and Michele Magno are with the Center for Project Based Learning, ETH Z{\"u}rich, Z{\"u}rich, Switzerland}
\thanks{Christoph Leitner and Luca Benini are with the Integrated Systems Laboratory, ETH Z{\"u}rich, Z{\"u}rich, Switzerland}
\thanks{Luca Benini is also with the Department of Electrical, Electronic, and Information Engineering, University of Bologna, Bologna, Italy}
\thanks{Corresponding author: christoph.leitner@iis.ee.ethz.ch}
}

\markboth{Journal of \LaTeX\ Class Files,~Vol.~18, No.~9, September~2020}
{How to Use the IEEEtran \LaTeX \ Templates}

\maketitle

\begin{acronym}
\acro{us}[US]{ultrasound}
\acro{tx}[Tx]{transmits}
\acro{rx}[Rx]{received}
\acro{ecg}[ECG]{electrocardiography}
\acro{ppg}[PPG]{photoplethysmography}
\acro{hr}[HR]{heart rate}
\acro{iot}[IoT]{Internet-of-Things}
\acro{fpga}[FPGA]{Field Programmable Gate Arrays}
\acro{cots}[COTS]{common off-the-shelf}
\acro{tof}[ToF]{Time-of-Flight}
\acro{fft}[FFT]{Fast Fourier Transform}
\acro{snr}[SNR]{signal-to-noise ratio}
\acro{2d}[2D]{Two Dimensional}
\acro{pdms}[PDMS]{Polydimethyl-Siloxane}
\acro{mcu}[MCU]{microcontroller}
\acro{pll}[PLL]{phase-locked loop}
\acro{dma}[DMA]{Direct memory access}
\acro{opamp}[op-amp]{operational amplifiers}
\acro{prt}[PRT]{Pulse-Repetition Time}
\acro{prf}[PRF]{pulse-repetition frequency}
\acro{spt}[SPT]{Sampling Time}
\acro{dsp}[DSP]{digital signal processing}
\acro{fifo}[FIFO]{First In, First Out}
\acro{soa}[SoA]{State-of-the-Art}
\acro{hmi}[HMI]{Human-Machine Interface}
\acro{led}[LED]{light-emitting diodes}
\acro{2d}[2D]{two-dimensional}
\acro{adc}[ADC]{analog-digital converters}
\acro{soc}[SoC]{system-on-chip}
\acro{ic}[IC]{integrated circuit}
\acro{rf}[RF]{radio frequency}
\acro{ic}[IC]{integrated circuit}

\end{acronym}
\begin{abstract}

This work explores the feasibility of employing \ac{us} technology in a wrist-worn \ac{iot} device for low-power, high-fidelity \ac{hr} extraction.
\ac{us} offers deep tissue penetration and can monitor pulsatile arterial blood flow in large vessels and the surrounding tissue, potentially improving robustness and accuracy compared to \ac{ppg}.

We present an \ac{iot} wearable system prototype utilizing a commercial \ac{mcu} employing the onboard \ac{adc}  to capture high frequency \ac{us} signals and an innovative low-power \ac{us} pulser. An envelope filter lowers the bandwidth of the \ac{us} signal by a factor of $>$\qty{5}{x}, reducing the system's acquisition requirements without compromising accuracy (correlation coefficient between \ac{hr} extracted from enveloped and raw signals, r(\qty{92}{})=\qty{0.99}{}, p$<$\qty{0.001}{}). The full signal processing pipeline is ported to fixed point arithmetic for increased energy efficiency and runs entirely onboard. The system has an average power consumption of \qty{5.8}{\milli\watt}, competitive with \ac{ppg} based systems, and the \ac{hr} extraction algorithm requires only \qty{68}{\kilo\byte} of RAM and \qty{71}{\milli\second} of processing time on an ARM Cortex-M4 based \ac{mcu}. The system is estimated to run continuously for more than \qty{7}{days} on a smartwatch battery.

To accurately evaluate the proposed circuit and algorithm and identify the anatomical location on the wrist with the highest accuracy for \ac{hr} extraction, we collected a dataset from \qty{10}{} healthy adults at three different wrist positions. The dataset comprises roughly \qty{5}{hours} of \ac{hr} data with an average of \qty{80.6}{}$\pm$\qty{16.3}{bpm}. During recording, we synchronized the established \ac{ecg} gold standard with our \ac{us}-based method. The comparisons yields a Pearson correlation coefficient of r(\qty{92}{})=\qty{0.99}{}, p$<$\qty{0.001}{} and a mean error of \qty{0.69}{}$\pm$\qty{1.99}{bpm} in the lateral wrist position near the radial artery.

The collected dataset and code used in this work have been open-sourced and are available at \href{https://github.com/mgiordy/Ultrasound-Heart-Rate}{https://github.com/mgiordy/Ultrasound-Heart-Rate}. 

\end{abstract}


\renewcommand{\arraystretch}{1.5}

\section{Introduction}
\label{sec:introduction}

\IEEEPARstart{U}{ltrasound} (\acsu{us}) technology finds applications in many areas, including medical imaging for diagnostic purposes \cite{ensminger_ultrasonics_2024,9691276}, non-destructive testing for material assessment \cite{silva2023}, object detection and ranging approaches \cite{fitzpatrick_airborne_2020}, or underwater communication \cite{schulthess_passive_2024}. In \ac{us}, pulse-echo is a fundamental operating mode in which a transducer \ac{tx} short acoustic pulses into the medium being examined \cite{szabo_diagnostic_2014}. These sound waves interact with medium boundaries, causing partial reflection of the acoustic energy. 
The analysis of \ac{rx} echoes in the resulting amplitude-time curves (A-mode) reveals information about the depth, mechanical properties, and even the functional states of the investigated medium \cite{jin_estimation_2024}. 
Moreover, an A-mode scan line can be repeatedly probed at a specific \ac{prf}, and the received signals are recorded and stacked over time. This time-motion mode (M-Mode) generates a \ac{2d} visualization, illustrating the temporal evolution of medium properties along the individual A-mode lines \cite{mohammadzadeh_asl_beamforming_2024}. Consequently, analyzing frequency behavior in both the A-mode channel's depth and its temporal development facilitates examining the medium's functional behavior. 
Alternatively, multiple channels can be used to create \ac{2d} grey-scale images (B-mode) based on the received echoes using beamforming algorithms \cite{mohammadzadeh_asl_beamforming_2024}. 

Specifically in medical applications, \ac{us} offers non-ionizing deep tissue penetration. Its safety and multiple modalities make \ac{us} the most versatile tool used in various medical analyses \cite{daniels_practical_2016}. For example, \ac{us} is employed to visualize internal structures such as organs \cite{wang2022} or the musculoskeletal system \cite{hu2023}, it provides insights into function and tissue morphology \cite{leitner2019}, and allows the identification of possible abnormalities \cite{kruse_muscle_2018}. Furthermore, \ac{us} is also a key technology for rapid assessment of blood flow, for example, to detect blockages or stenosis in blood vessels \cite{kenny2021} that can restrict flow and increase the risk of stroke and other diseases. Medical \ac{us} applications today extend beyond traditional clinical settings \cite{yang2024, 9691276}. Recent advances in materials, electronics, and algorithms have facilitated the miniaturization of transducers \cite{9691276, wang_continuous_2021, wang2022, hu2023, du_conformable_2023, keller2023}, the development of smart, wearable \ac{us} readout electronics \cite{frey_wulpus_2022} and completely integrated wearable \ac{us} systems (including transducers and acquisition electronics) \cite{lin2024}. This has unlocked possibilities for \ac{us} applications in previously unexplored areas, including the \ac{iot}\cite{wang2019, liu2023, 9219124}, augmented and extended reality \cite{mcintosh2017} as well as continuous vital sign \cite{lin2024} and cardio-respiratory monitoring \cite{vostrikov_complete_2023}. Despite the promising potential of \ac{us} technology in wearable devices, several challenges remain. Minimizing power consumption, ideally to a level supported by a few hundred \qty{}{\milli\ampere\hour} battery for all-day operation \cite{AppleWat86:online}, necessitates efficient on-device signal processing algorithms \cite{frey_wearable_2023}. Additionally, optimizing transducer placement and tissue coupling can enhance the robustness of this approach \cite{lin2024}. Addressing these challenges is crucial to fully harness the potential of wearable ultrasound-based vital sign monitoring solutions.

In clinical settings, electrocardiography (ECG) remains the gold standard for high-resolution measurement of the heart's activity \cite{goldberger_front_2018}. However, the introduction of single-lead \ac{ecg} capabilities in smartwatches like the Apple Watch Series 4 (introduced in 2018) \cite{apple_take_nodate} has opened doors for consumer-based heart rhythm monitoring. Studies have shown promising accuracy using these devices for \ac{ecg} measurements \cite{spaccarotella_measurement_2021}, and other major brands have rapidly adopted the technology \cite{fitbit_fitbit_nodate, samsung_measure_nodate, garmin_ecg_nodate, polar_polar_nodate}. Despite its accuracy, single-lead \ac{ecg} on smartwatches is not a perfect replacement for \ac{hr} monitoring on the go, as there are limitations regarding posture and physical contact points. For example, for most smartwatches, users are required to remain still and maintain contact with a wrist electrode (frequently placed underneath the display) and a finger electrode (usually situated on the crown or the screen) during measurements \cite{apple_take_nodate}. Therefore, most smartwatches have additional optical \ac{ppg} sensors to enable continuous \ac{hr} detection. \ac{ppg} utilizes \ac{led} to shine light onto the skin. Different tissues and arterial and venous blood absorb light with certain wavelengths at varying degrees \cite{allen_photoplethysmography_2022}. By measuring changes in reflected light intensity using a photodiode receiver, \ac{ppg} sensors can estimate blood volume variations, which correlate with \ac{hr} \cite{de_pinho_ferreira_review_2021}. To achieve sufficient \ac{rx} signal strength, however, \ac{ppg} sensors must operate \ac{led}s with sufficient brightness and output current, resulting in power envelopes in the range of tens of milliwatts \cite{ebrahimi_ultralow-power_2023}. Thus, to minimize sensor power consumption, \ac{led}s are usually aggressively duty-cycled or operated in an event-based fashion \cite{ebrahimi_ultralow-power_2023, apple_monitor_nodate}. 

Despite the availability of \ac{cots} low-power \ac{ic} that make \ac{ppg} sensors an attractive option \cite{analog_devices_max86150_nodate, texas_instuments_afe4900_nodate}, a variety of factors can significantly impact their performance \cite{ray_review_2023}. While both \ac{ppg} and \ac{us} can be affected by common motion artefacts or sensor detachment, key limitations unique to PPG are interference from ambient light and restricted penetration depths \cite{biswas_heart_2019}. Specifically \ac{ppg} signal quality suffers from:
\begin{inparaenum}[(i)]
\item the predominance of a direct current (DC) component compared to the low pulsating alternating current (AC) component in the \ac{ppg} signal collected at the photodiode and the necessary amplification and noise suppression \cite{loh_application_2022}, 
\item the interplay of varying optical properties in tissues and blood, influenced by factors like skin colour \cite{koerber2023accuracy} or the presence of hair and tattoos \cite{nelson2020guidelines}, and
\item the chosen wavelength (e.g., red, green or infrared) used to illuminate tissues and their corresponding penetration depths and varying forward voltages \cite{long_wearable_2022}.
\end{inparaenum}
Conversely, pulse-echo \ac{us} offers deep tissue access and the ability to directly monitor pulsatile arterial blood flow in large vessels and their surrounding tissues, e.g., the radial or ulnar arteries in the wrist.

We investigate the potential of low-power \ac{us} for \ac{hr} extraction from a single \ac{us} transducer and its optimal placement on the wrist, enabling accurate long-lasting wearable \ac{us} \ac{iot} Devices. We exploit an analog envelope circuit to reduce the bandwidth of the \ac{us} signal, enabling its acquisition and processing by a low-power ARM Cortex-M4 \ac{mcu} suitable as the core of a long-lasting wearable IoT. We show a proof-of-concept wearable \ac{us} system for high-accuracy \ac{hr} extraction, analyzing its power consumption, computational demands, and memory requirements.
We validated our system and algorithm on data collected from 10 healthy subjects and demonstrated that low-power and energy-efficient \ac{us} can achieve ECG-level accuracy on \ac{hr} extraction. In particular, the major contributions of this work are:

\begin{itemize}
        \item Investigating the accuracy of \ac{hr} extraction employing pulse-echo \ac{us} and its positioning on the wrist, achieving a correlation coefficient of r(\qty{92}{})=\qty{0.99}{}, p$<$\qty{0.001}{} between ECG and \ac{us} with an error of \qty{0.69}{}$\pm$\qty{1.99}{bpm} for the lateral wrist position. This level of accuracy is competitive with that of commercially available wrist-worn \ac{ppg} devices \cite{biswas_heart_2019, hahnen_accuracy_2020, sarhaddi_comprehensive_2022, hajj-boutros_wrist-worn_2023}, suggesting the potential of our approach for accurate and robust \ac{hr} monitoring,
        \item Employing a frequency reduction technique to lower the bandwidth of the \ac{us} frequency by a factor of $>$5x without compromising \ac{hr} estimation accuracy,
        \item Design and development of a signal processing algorithm in fixed-point arithmetic for energy-efficient hardware usage,
        \item Conducting full memory, latency, and power profiling for the deployment in a potential wearable device, resulting in \qty{68}{\kilo\byte} memory footprint, \qty{71}{\milli\second} \ac{dsp} processing time, and \qty{5.8}{\milli\watt} total power consumption (including the transmit (TX) and receive (RX) functionalities of \ac{us}, the analog frontend and the \ac{dsp}),
        \item Open-sourcing the dataset, as well as the code used in this work in the hope of fostering further development in wearable \ac{us}\footnote{https://github.com/mgiordy/Ultrasound-Heart-Rate}.
\end{itemize}

This paper is organized as follows. Section~\ref{sec:related_work} reviews existing research on piezoelectric wrist-based vital sign monitoring and wearable \ac{us} circuit architectures. Section~\ref{sec:preliminaries} provides background information on \ac{us} data handling and \ac{hr} extraction algorithms. Section~\ref{sec:dataset} describes the methodology used for data collection, including the experimental setup and the employed protocol. Section~\ref{sec:methods} delves into the hardware components utilized for the embedded implementation and details the efficient implementation of the \ac{hr} extraction algorithm. Section~\ref{sec:results} presents the accuracy of the HR measurements and the profiling of the ported algorithm and sets these findings in perspective. Section~\ref{sec:limitations} acknowledges the current limitations of our work and identifies potential future research directions in wearable \ac{us} technology for wrist-based applications. Finally, Section~\ref{sec:conclusion} summarizes this work's key findings and contributions, concluding the paper.
\begin{table*}[th!]
    \centering
    \setlength{\tabcolsep}{3pt}
    \begin{center}
    \begin{tabular}{ccccccccc}
    \toprule
    Study   
    & Modality    
    & \begin{tabular}[c]{@{}c@{}}Transd.\\Type\end{tabular}
    & \begin{tabular}[c]{@{}c@{}}C. Freq.\\$[$MHz$]$\end{tabular}   
    & \begin{tabular}[c]{@{}c@{}}Acoustic\\Coupling\end{tabular}   
    &  \begin{tabular}[c]{@{}c@{}}Integration\\Level\end{tabular}    
    &\begin{tabular}[c]{@{}c@{}}Study\\Participants\end{tabular}  
    & \begin{tabular}[c]{@{}c@{}}Error\\Metric\end{tabular}\\
    \midrule

        \cite{fang_wrist_2021} 
        &Rx       
        &Polypropylene
        &N.A.
        &Direct Contact    
        &Lab Bench
        &7  
        &N.A.\\

        \cite{polley_wearable_2021}
        &Rx     
        &Piezo Ceramic \cite{pui_audio_benders_nodate}
        &0.009
        &Silicone \cite{barnes_pinkysil_nodate}
        &Board Level
        &3
        &-1.8 bpm$^\dagger$ \\

        \cite{peng_noninvasive_2021}
        &TxRx  
        &\begin{tabular}[c]{@{}c@{}}PZT-5A/PDMS\\1-3 Anisotropic Comp.\end{tabular}
        &4.73
        &Liquid Gel
        &Lab Bench
        &1
        &-1 bpm$^{\dagger\dagger}$\\

        \cite{park_opto-ultrasound_2022}
        &TxRx
        &Lithium Niobate
        &6
        &Parylene C
        &Lab Bench
        &$>$1$^\ddagger$
        &in agreement$^{\ddagger\ddagger}$\\

        \cite{shumba_monitoring_2024}
        &Rx
        &Aluminium Nitride
        &N.A.
        &Resin \cite{formlabs_elastic_nodate}
        &Lab Bench
        &2
        &0.29 bpm$^\|$\\

    \midrule
        \addlinespace[4pt]
        \textbf{This work}
        &\textbf{TxRx}
        &Piezo Ceramic \cite{piezo_hannas_customize_nodate}
        &10
        &Gel Pad
        &\textbf{Board Level}
        &10
        &\textbf{-0.68 (1.95)} \\
    \bottomrule
    \end{tabular}
    \end{center}

    \begin{tablenotes}
        \item $^\dagger$ stated error between \ac{ecg} and proposed method (additionally $R^2$ was given with 0.99), $^{\dagger\dagger}$ the \ac{hr} extracted by the upper arm blood pressure monitor was 71 bpm whereas the \ac{hr} extracted with their device was 70 bpm, $^\ddagger$ participants are given in plural so we assume that it is more than 1, $^{\ddagger\ddagger}$ authors state that their \ac{ppg} and \ac{us} sensors are in agreement with each other, $^\|$ the indicated error after the outlier removal is 0.19 bpm.
    \end{tablenotes}
    \vspace{0.3cm}
    \caption{This table compares acoustic wrist-worn \ac{hr} measurement methods. Studies are categorized based on the sensor functionality: microphone-only (only \ac{rx}) or pulse-echo mode operation (\ac{tx} and \ac{rx}). Additionally, the table specifies the prototype stage: A physical proof of concept (utilizes lab equipment wired together for operation and demonstrates the core functionality), a device with board-level integration (features all necessary components soldered onto a printed circuit board). The table additionally includes center frequencies of the transducers (where available) and indications of acoustic coupling between the transducer and the skin. It further details the number of participants in each study and the accuracy of the \ac{hr} detection.}
    \label{tab:realtedwork}
    \vspace{-0.3cm}
\end{table*}

\section{Related Work}
\label{sec:related_work}

\subsection{Acoustic Wrist-Based HR Detection}
Smartwatches have evolved beyond their timekeeping origins, occupying the wrist as a prime location for on-the-go health monitoring and fitness tracking \cite{henriksen_using_2018, islam_visualizing_2020}. This evolution goes beyond the traditional functions of a watch, offering a diverse range of functionalities, including heart rate monitoring \cite{nelson_guidelines_2020}, gesture recognition \cite{jiang_emerging_2022} or behavioural analyses \cite{suh_worker_2024}. In that context, acoustic techniques are being explored as alternative modalities for \ac{hr} monitoring on the wrist \cite{de_pinho_ferreira_review_2021}. Table \ref{tab:realtedwork} summarizes existing research on these approaches for wrist-based \ac{hr} assessment. Over half of the reviewed studies employ piezoelectric transducers as passive microphone receivers within their systems. While this approach simplifies circuit design by requiring only low-frequency \ac{rx} stages, it fails to leverage the full potential of piezoelectrics. These excel at generating high-frequency mechanical wave pulses (\qty{1}{} - \qty{20}{\mega\hertz}) that can penetrate deep tissues and facilitate the detection of pulsatile signals originating from deeper-lying and larger arterial structures \cite{lin2024}.   
Conversely, Peng et al. \cite{peng_noninvasive_2021}, and Park et al. \cite{park_opto-ultrasound_2022} showcase the application of pulse-echo \ac{us} for \ac{hr} detection on the wrist. They use transducers with center frequencies of \qty{4.73}{} and \qty{6}{\mega\hertz}, respectively. However, their studies are limited to controlled laboratory settings, which lack complexities encountered with miniaturization and the edge deployment of algorithms (i.e., constraints in power consumption, data rates, etc.).  

Notably, a common trend in achieving acoustic coupling in almost all studies involved using polymer materials or hydrogel-based interfaces. This preference aligns with typical materials used for watch bands such as silicons or rubbers \cite{swatch_replacement_nodate} and could facilitate seamless integration of transducers. In addition, the use of polymer transducers \cite{fang_wrist_2021} 
could even further enhance this integration process due to their impedance properties \cite{keller2023}, often closely resembling those of human tissue.

Moreover, we found that most studies did not show significant sample sizes and that statistics and error metrics were inconsistent or not reported at all, albeit guidelines have been proposed \cite{nelson_guidelines_2020}. 

In contrast to previous research, this study presents the first wrist-worn \ac{hr} monitor incorporating a high-frequency (\qty{10}{\mega\hertz}) piezoelectric transducer for deep tissue penetration. The system is implemented through a board-level embedded design optimized for wearable applications, addressing the inherent challenges associated with such integration. In addition, we collected data from more participants than any other previous study, strengthening the validity of our approach.

\begin{figure*}[thb]
    \centering
    \begin{minipage}{0.23\textwidth}
        \centering
        \includegraphics[width=\linewidth]{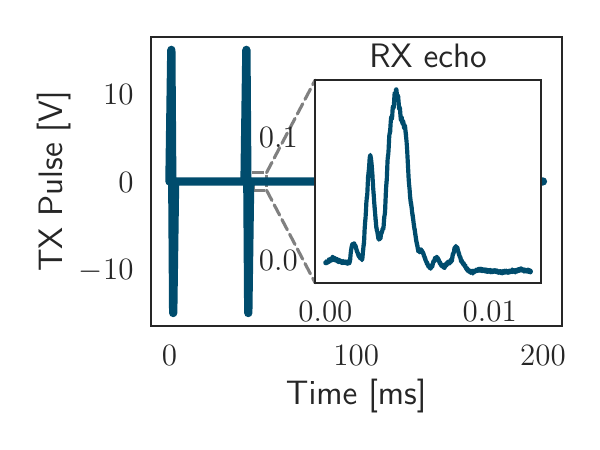}
        \subcaption{}
    \end{minipage}
    \hfill
    \begin{minipage}{0.23\textwidth}
        \centering
        \includegraphics[width=\linewidth]{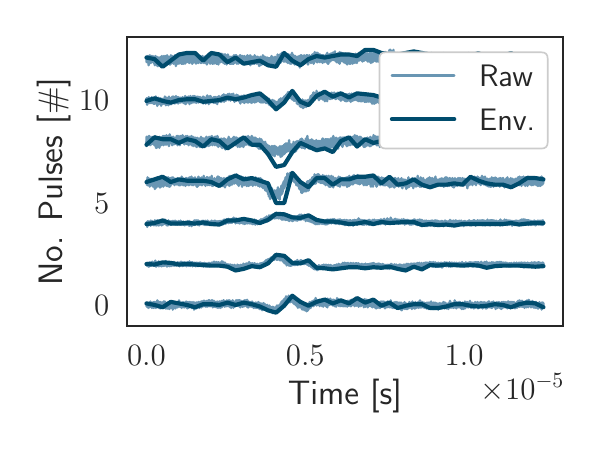}
        \subcaption{}
    \end{minipage}
    \hfill
    \begin{minipage}{0.23\textwidth}
        \centering
        \includegraphics[width=\linewidth]{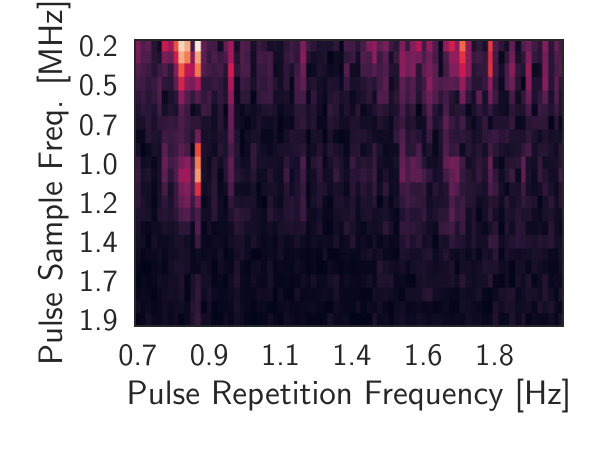}
        \subcaption{}
    \end{minipage}
    \hfill
    \begin{minipage}{0.23\textwidth}
        \centering
        \includegraphics[width=\linewidth]{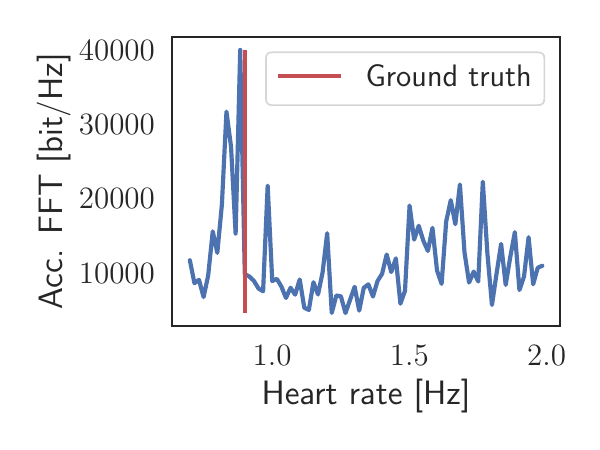}
        \subcaption{}
    \end{minipage}
    \caption{
    The figure depicts the chosen signal processing pipeline for \ac{hr} extraction on the wrist. Panel (a) visualizes the transmitted \ac{us} pulses and corresponding enveloped echoes. Panel (b) displays the time-differentiated signals. Panel (c) shows the application of a 2D FFT to the differentiated signals. Finally, panel (d) illustrates the frequency accumulation from the 2D FFT and the identified peak corresponding to the \ac{hr}.}
    \label{fig:signal_processing}
\end{figure*}

\subsection{Wearable US Circuit Architectures}
Recent technological advances have driven the development of wearable \ac{us} systems, particularly for monitoring vital signs using patch-type devices \cite{lin2024}. Despite these latest advances, one of the biggest challenges in integrating \ac{us} into future wearable systems remains their power consumption: a critical factor for wearable technology, where battery life is key \cite{benini_wireless_2006}. As emphasized in \cite{vostrikov_towards_2021}, factors such as an increased number of \ac{rx} channels, the selection of \ac{us} frequencies, and \ac{prf} contribute significantly to the power consumption of \ac{us} systems. To meet all these requirements in the context of \ac{us}, advanced strategies are needed to manage the increasing data volume and the associated overhead for conversion, storage, and transmission.

\begin{table}[t]
\centering
\setlength{\tabcolsep}{3pt}
\begin{center}
\begin{tabular}{cccccc}
    \toprule
    Study 
    & \begin{tabular}[c]{@{}c@{}}Power\\$[$mW$]$\end{tabular}
    & \# Ch.
    &  \begin{tabular}[c]{@{}c@{}}BW\\Red.\end{tabular}   
    & \begin{tabular}[c]{@{}c@{}}Embedded\\Platform\end{tabular}
    & \begin{tabular}[c]{@{}c@{}}Feature\\Extraction\end{tabular}\\

    \midrule
    WMAUS \cite{yang2019wearable}
    & 3500   
    & 8
    & \xmark
    & \begin{tabular}[c]{@{}c@{}}DSP$^\dagger$\\dsPIC33\end{tabular}
    & Offline$^\ast$\\ 
    
    Song et al.\cite{song2019design}           
    & 3700
    & 2
    & \cmark
    & \begin{tabular}[c]{@{}c@{}}FPGA$^{\dagger\dagger}$\\Xilinx Spartan 6\end{tabular}
    & Offline$^\ast$\\

    MoUsE \cite{fournelle2021portable}    
    & 12000   
    & 32
    & \xmark
    & \begin{tabular}[c]{@{}c@{}}FPGA\\Xilinx ZYNQ 7\end{tabular}
    & Offline\\

    Yin et al.\cite{yin2022wearable}        
    & $\sim$ 1600
    & 1$^\|$
    & \xmark
    & \begin{tabular}[c]{@{}c@{}}MCU$^{\ddagger}$\\STM32 F7\end{tabular}
    & Offline$^\ast$\\

    WULPUS \cite{frey_wulpus_2022} 
    & \textless{}25
    & 1$^\mathsection$
    &\xmark
    & \begin{tabular}[c]{@{}c@{}}MCU\\MSP430FR\end{tabular}
    & Offline$^\ast$\\
    
    USoP \cite{lin2024}         
    & 614   
    & 1$^{\mathparagraph}$
    &\xmark
    & \begin{tabular}[c]{@{}c@{}}MCU\\PIC32MZ EF\end{tabular}
    & Offline$^\ast$\\
    
    \midrule
    
    This work 
    & 5.8 
    & 1
    &\cmark
    & \begin{tabular}[c]{@{}c@{}}MCU\\STM32L4\end{tabular}
    & On-device\\
    
    \bottomrule
\end{tabular}
\end{center}

\begin{tablenotes}
    \item 4$^\|$, 8$^\mathsection$, 32$^{\mathparagraph}$, channels time-multiplexed, $^\dagger$digital signal processor, $^{\dagger\dagger}$field programmable gate array, $^{\ddagger}$microcontroller
    \item $^\ast$wireless link
\end{tablenotes}

\caption{This table compares state-of-the-art, low-power wearable \ac{us} devices. It details their power consumption, number of channels used, consideration of bandwidth reduction techniques, core platform (MCU or FPGA), and whether online feature extraction was implemented.}
\label{tab:device_comparison}

\end{table}

Recent research \cite{yin2022wearable,frey_wulpus_2022,lin2024} in wearable \ac{us} electronics show a shift towards time-multiplexed single channel devices incorporating \ac{mcu} with enhanced analog capabilities and build-in high performance \ac{adc} \cite{st_microelectronics_stm32l4_nodate, microchip_pic32mz1024efh064_nodate, texas_instuments_msp430fr5043_nodate}, shifting away from previously dominant \ac{fpga} in \ac{us} system architectures \cite{boni_ultrasound_2018}. Our work inserts itself in this transition, which is driven by the need for highly efficient edge computing and the tight integration of mixed-signal functionalities very close to the actual sensing site (transducer) and can be seen as a trend in \ac{iot} in general \cite{kong_edge_2022}. Requirements for these \ac{mcu}s are also significantly impacted by the choice of \ac{us} transducers. In particular, their centre frequencies and bandwidths affect the minimum required sampling rates of \ac{adc}s. This is due to the adherence to the Nyquist-Shannon sampling theorem, which necessitates a sampling rate exceeding the Nyquist frequency. This interplay impacts electronic designs and transducer selection and suggests that a co-design, as demonstrated in \cite{lin2024, giordano_towards_2023}, can be an efficient way to balance the trade-offs in wearable \ac{us} hardware designs.

As an example, Frey et al. \cite{frey_wulpus_2022} prioritize low power consumption ($<$ \qty{25}{\milli\watt}) and operate their device up to maximum \ac{us} frequencies of $\sim$\qty{4}{\mega\hertz}. They achieve this with an internal \ac{adc} working at \qty{8}{msps}$@$\qty{12}{\bit} resolution on a MSP430FR \cite{texas_instuments_msp430fr5043_nodate} \ac{mcu} from Texas Instruments. Lin et al. \cite{lin2024} on the other hand prioritize achieving higher \ac{us} frequencies (up to \qty{6}{\mega\hertz}) and utilize a PIC32MZ \cite{microchip_pic32mz1024efh064_nodate} \ac{mcu} from Microchip. Their on-board \ac{adc} boasts a
sampling frequency of \qty{12.5}{msps}@12bit resolution at the cost of increased power consumption (\qty{614}{\milli\watt}), two orders of magnitude higher than our approach. Moreover, it is essential to note that all newer \ac{mcu} based devices operate in single-channel \ac{rx} mode. Hence, to accommodate larger transducer arrays, time resolutions are sacrificed through time-multiplexing to stay within wearable system power constraints.

Interestingly, apart from \cite{song2019design} that employ quadrature detection, none of the reviewed devices utilizes bandwidth reduction techniques to either accommodate transducers with higher centre frequencies and thus higher resolutions or to further relax design constraints in existing devices. Moreover, current wearable platforms lack the ability to perform advanced \ac{us} \ac{rf} data processing directly on the edge. Instead, they function as standalone units, collecting \ac{rf} data and transmitting it wirelessly to more powerful servers for computation and display. However, these server-centric approaches depend on reliable high-speed wireless connections and can restrict user mobility and accessibility. Additionally, latency issues can hinder user experience for applications requiring real-time responses \cite{merenda_edge_2020}.

In contrast to previous research, we present the first \ac{mcu}-based \ac{us} system architecture incorporating an envelope detection circuit in the \ac{rx} path. This novel design approach enables the utilization of high-frequency, high-resolution transducers, e.g., as recently proposed flexible, polymer thin-film interfaces \cite{keller2023}, while simultaneously accommodating a low-frequency \ac{adc} operating at \qty{4}{msps}$@$\qty{12}{\bit} resolution, as typically found in conventional mixed-signal \ac{mcu}s. Furthermore, to the best of our knowledge, we are the sole contributors of an on-device feature extraction algorithm, enabling substantial power savings compared to previous \ac{mcu} based \ac{us} systems demonstrated in the literature. 
\begin{figure*}[thb]
    \centering
    \includegraphics[width=0.9\linewidth,trim=0 0 0 0cm,clip]{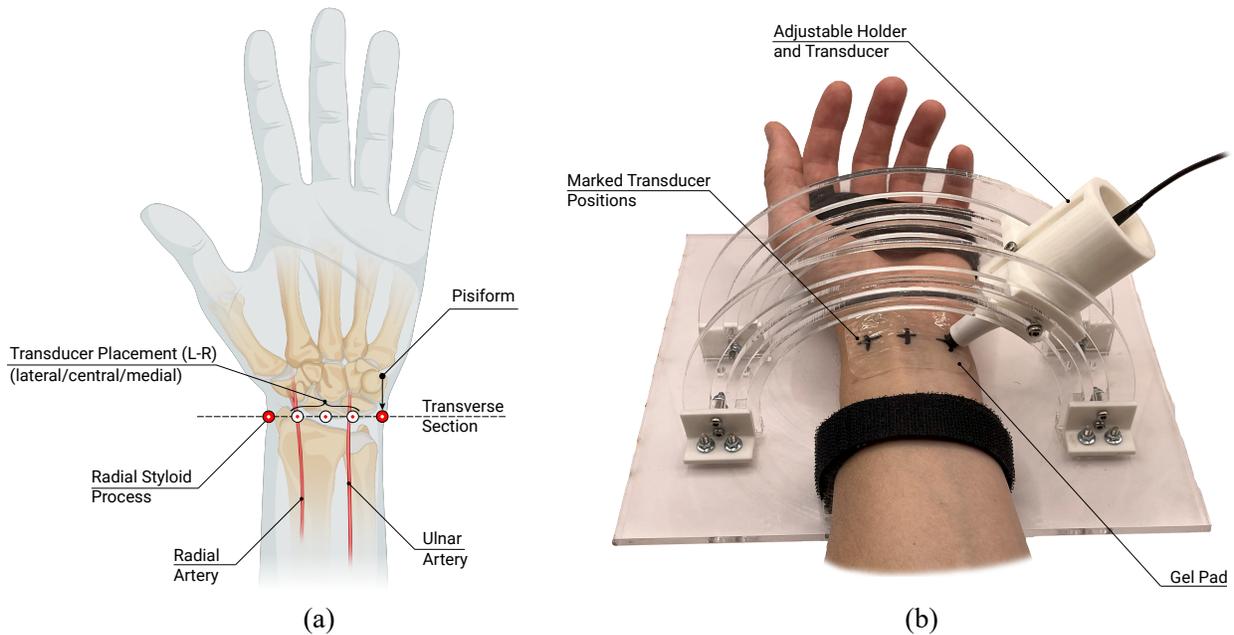}
    \caption{Panel (a) identifies three equidistant anatomical landmarks on the wrist suitable for wearable smart wristband transducer placement. These points lie between the radial styloid process and the projected location of the pisiform bone onto the cross-section passing through the radial styloid process. Panel (b) shows the adjustable test bench designed and fabricated using open-sourced 3D-printed and laser-cut parts. This bench features a polar plane that allows for transducer rotation around the wrist and radial position adjustment to optimize skin contact.}
    \label{fig:wrist_setup}
\end{figure*}

\section{Preliminaries}
\label{sec:preliminaries}

In the study of tissue dynamics, differentiation is a commonly used technique to isolate the time-varying components of a signal \cite{waasdorp_combining_2021}. Applying a \ac{2d} \ac{fft} to M-mode matrices differentiated along the sampling direction allows the temporal variations to be effectively decomposed into their constituent frequency components, revealing any underlying periodicities within the signals. To extract \ac{hr} from M-mode pulse-echo \ac{us} signals obtained from the wrist we leverage the \ac{2d} \ac{fft}, adapting an algorithm proposed for cardio-respiratory monitoring \cite{vostrikov_complete_2023}. This algorithm was further refined by the authors in \cite{giordano_towards_2023} for enveloped \ac{us} signals. Figure~\ref{fig:signal_processing} visualizes the processing steps applied to the signal, which can be summarized as follows:
\begin{itemize}
    \item \textbf{M-Mode representation:} the acquired A-Mode data, shown in Figure \ref{fig:signal_processing}(a) are put in a matrix, where each \ac{us} echo occupies one row. This aligns the \ac{spt} along the horizontal axis and the \ac{prt} along the vertical axis.
    \item \textbf{Differentiation:} to enhance the \ac{hr} signal encoded in the M-Mode matrix, the velocity of the signal is computed by differentiating along the \ac{prt} axes. The output of this step can be seen in Figure \ref{fig:signal_processing}(b).
    \item \textbf{2D \ac{fft}:} to extract the \ac{hr} we apply two \ac{fft}s, along the \ac{spt} and the \ac{prt}, moving in the frequency domain. The output of this operation can be seen in Figure \ref{fig:signal_processing}(c).
    \item \textbf{Frequency accumulation:} the absolute values of the \ac{fft} run on the \ac{spt} axis are accumulated along the \ac{prt} axis. With this step, we highlight the signal's time-periodic behaviours and reduce the dimensionality from 2D to 1D. This operation is graphed in \ref{fig:signal_processing}(d).
    \item \textbf{Peak detection:} From the last step, we already have a signal in the frequency domain on the \ac{prt} axis, which relates to periodic behaviours in time. Therefore, the last step of the signal processing pipeline involves finding the peak in a range of frequencies that make sense for \ac{hr} detection; in this work, we choose the interval [\qty{0.5}{\hertz}, \qty{2}{\hertz}], corresponding to [\qty{30}{bpm}, \qty{120}{bpm}], as it encompasses the typical resting and light physical activity heart rates \cite{canning_individuals_2014}.
\end{itemize}
\begin{table}[htb]
    \centering
    \begin{tabular}{ccc}
        \toprule
        Participants 
        & Avg. Age [years]
        & Avg. Heart Rate [bpm]\\
        \midrule
        10 male
        & $28.4 \pm 6.3$
        & 80.1 $\pm$ 16.3 \\
        \bottomrule
    \end{tabular}
    \caption{Structure of the data set collected from synchronized \ac{ecg} and \ac{us} signals. }
    \label{tab:dataset}
\end{table}

\section{Dataset Collection}
\label{sec:dataset}

In our previous work \cite{giordano_towards_2023}, we demonstrated the identification of the pulsation frequencies of a peristaltic pump using A-mode \ac{us} on a phantom. This study extends their work by aiming to validate these results in a real human environment on the wrist. We collected a data set on ten healthy adults (Table \ref{tab:dataset}) to develop our embedded algorithm and optimize the component design for our wearable \ac{hr} monitoring system using \ac{us}. Moreover, we specifically designed the experimental setup and protocol to address two key research questions:
\begin{enumerate}[i]
    \item Can \ac{hr} be accurately extracted from a single \ac{us} transducer on the human wrist? We aim to streamline the setup by utilizing a single \ac{us} transducer, thus minimizing the volume of data collected and the overall power consumption for the acquisition and processing. Consequently, this approach facilitates a more wearable deployment. 
    \item What is the optimal placement of the \ac{us} transducer on the wrist to achieve the highest \ac{snr}? Determining the influence of different placements on the \ac{snr} is crucial for a reliable and robust extraction of \ac{hr} from \ac{us}. Furthermore, it can optimize future wristband designs incorporating \ac{us}.
\end{enumerate}

Synchronized ECG recordings were acquired from the chest and coupled with single-channel \ac{us} acquisitions at three distinct wrist locations (Figure~\ref{fig:wrist_setup} (A)). We collected 300 one-minute recordings with \ac{hr} ranging from 48-\qty{128}{bpm}. Six recordings were excluded from the study: one due to missing data and five others because the \ac{us} transducer detached from the wrist. Data was collected during rest and at regular intervals following light physical activity. Recording data immediately after exercise and during the cool-down period allowed us to capture a broader range of \ac{hr}s, thus strengthening the assessment of the system's robustness. The study was approved by the local University's Ethics Committee (registration number: EK 2024-N-200), and written informed consent was obtained from all participants.

\subsection{Positioning On The Wrist}
\label{subsec:positions_on_the_wrist}
The radial and ulnar arteries form a terminal branched network supplying blood to the wrist and hand, as graphed in Figure \ref{fig:wrist_setup} (A). Although the ulnar artery can be palpated on the anterior and medial side of the wrist, the radial artery lies more superficially around the pronator quadratus muscle and is only covered by a thin layer of fascia and skin, hence it is usually more accessible and the pulse is more pronounced.

To find the best-performing position in wrist-worn applications like smartwatches or fitness trackers, we evaluated the transducer placement at three locations on the proximal aspects of the wrist. To find the best-performing position in wrist-worn applications like smartwatches or fitness trackers, we evaluated the transducer placement at three locations on the proximal aspects of the wrist. To ensure consistent measurement across subjects while accounting for individual anatomical variations, we identified two key landmarks on the wrist: the radial styloid process and the pisiform (Figure~\ref{fig:wrist_setup} (A)). We projected the location of the pisiform onto the cross-section passing through the radial styloid process. This projection point's distance from the radial styloid process was divided into four equal segments. The transducer was subsequently placed on the first (medial), second (central), and third (lateral) intersection point between the segments.

\subsection{Adaptable Wrist Measurement Test Bench}
\label{subsec:data_acquisition_device}

To ensure reproducibility and accommodate variations in participants' anatomy, we designed a custom data acquisition test bench. Figure \ref{fig:wrist_setup} (B) shows the device in use while collecting data from a study participant. The device has a rigid frame of 5 mm thick, laser-cut acrylic glass. This frame features two slots that allow a transducer holder to slide sideways (in a transverse plane) for medial/lateral adjustments. The holder can also be raised or lowered to accommodate different wrist sizes. Two straps complete the device, securing the participant's wrist and hand comfortably in distal/proximal directions. The device encompasses a polar plane centred on the subject's wrist, accurately targeting the desired point while accounting for variations in wrist shapes.

\subsection{Experimental Protocol}
\label{subsec:protocol}

Data was collected from 10 healthy volunteers.
We collected five 1-minute samples for each of the three positions on the wrist while resting and just after light exercise. Providing the full spectrum between resting and exercising \ac{hr}s, we test the system's robustness concerning diverse and varying \ac{hr}s.

\begin{figure*}[!tbp]
    \centering
    \includegraphics[width=\textwidth,trim=0 0 0 0cm,clip]{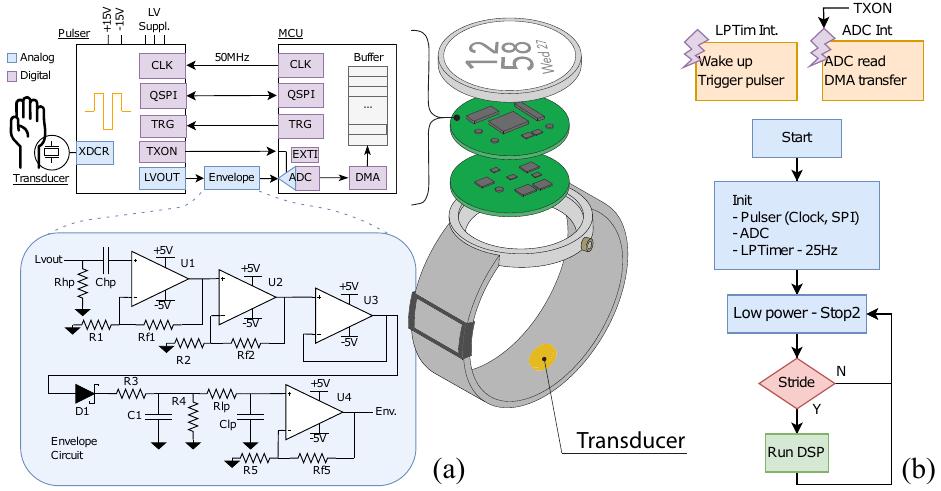}
    \caption{In (a), the electronic system overview is shown: the pulser is depicted in the top left corner, the \ac{mcu} is drawn in the top right, and the envelope filter's schematic is shown in the exploded view in the bottom. In (b), the firmware flowchart is detailed: the system defaults to low power mode after initialization. The LPTimer generates an interrupt and triggers a pulse, which in turn generates an ADC acquisition. If the configured stride is reached, then the \ac{dsp} is run, and \ac{hr} is computed.} 
    \label{fig:schematic_operation}
\end{figure*}

The following steps have been executed for each participant:
\begin{enumerate}
    \item Anatomical landmarks were identified on the wrist of the participant, and the transducer positions were marked on the skin as described in Section \ref{subsec:positions_on_the_wrist}.
    \item A gel pad was applied to the participant's wrist to ensure good acoustic coupling between the transducer and the skin. Gel pads were preferred to liquid gel, generally used for \ac{us} imaging, as they are deemed more suitable for the intended end application scenario. Moreover, the functionality of gel pads - i.e. the acoustic coupling - could ultimately be mimicked by a wristband made of soft, rubber-like material such as \ac{pdms} \cite{la_flexible_2022}.
    \item The participants were then asked to sit relaxed on a chair and insert their left hand inside the adaptable wrist measurement test bench (Figure \ref{fig:wrist_setup} (B)). The arm and hand were then secured by straps, and the transducer head was moved to match the positions marked before on the wrist.
    \item Fifteen minutes of \ac{us} data were collected in 1-minute intervals for five repetitions from three different wrist positions on each subject in a resting state.
    \item Following this, participants completed one minute of light exercise (jumping on the stand) to elevate their \ac{hr}. After the exercise, the participant returned to a seated position. The \ac{us} transducer was repositioned, and data acquisition continued for five minutes at one-minute intervals. This process was repeated for the remaining two transducer positions on the wrist, resulting in an additional 15 minutes of \ac{us} data per participant.
\end{enumerate}

\subsection{Data Collection Setup}

 To facilitate future research in the field of wearable \ac{us} devices, we designed our measurements to go beyond the data required for \ac{hr} detection. The data collected, and the sensors used to acquire our comprehensive dataset are described in the following paragraphs:

\subsubsection{Pulser}
The need for a configurable and low-power pulser drove the selection. The pulser selected was the \texttt{STHVUP32} (ST Microelectronics, Switzerland), set to drive 5 square pulses at \qty{10}{\mega\hertz} with a pulse repetition frequency of \qty{25}{\hertz}. The amplitude of the pulses was set to \qty{\pm15}{\volt}.

\subsubsection{Transducer}
For this study, we opted for a single-element \ac{us} transducer (Piezo Hannas, China) with a central frequency of \qty{10}{\mega\hertz}. This frequency selection aligns with the operating frequencies of emerging flexible \ac{us} sensors, as demonstrated in \cite{keller2023, giordano_towards_2023}.

\subsubsection{Data acquisition}
To acquire the \ac{us} data, we chose the \texttt{Teledyne ADQ412} ADC card (Teledyne Inc., USA). The acquisition board was connected to a PC via a PCI interface, and the data was acquired using a custom Matlab (MathWorks Inc., USA) script based on a set of APIs provided by the manufacturer. We acquired the raw \ac{us} echos as well as the analog enveloped \ac{us} signals (envelope filtered echos) at a sampling rate of \qty{2}{\giga\hertz} and a resolution of \qty{12}{\bit}. 

\subsubsection{Ground truth}
For ground truth reference, we employed a \texttt{Polar H10} chest belt (Polar Electro Oy, Finland). The belt transmitted \ac{hr} data via Bluetooth to the same PC that collected the \ac{us} signals. Both datasets were synchronized through a single Matlab script, ensuring precise temporal alignment. Additionally, the subject's \ac{ecg} was captured, providing data for potential future research in wearable \ac{us}-based cardiovascular monitoring.
The \ac{hr} of the participant was collected at a rate of \qty{1}{\hertz}, and the ECG was collected at a rate of \qty{104}{\hertz}.

\subsection{Data Analyses}
The agreement between the \ac{hr} derived from the enveloped \ac{us} signals and the reference \ac{ecg} recordings was assessed using Bland-Altman plots, and the Pearson correlation coefficient was computed to evaluate the linear relationship between \ac{hr} measurements. Correlations plots were employed to compare the \ac{hr} obtained from the enveloped \ac{us} signals with the \ac{ecg} data, as well as between \ac{hr} derived from the enveloped and raw \ac{us} signals.

\begin{figure*}
    \centering
    \includegraphics[width=1\textwidth,trim=0.8cm 0cm 0.8cm 0cm,clip]{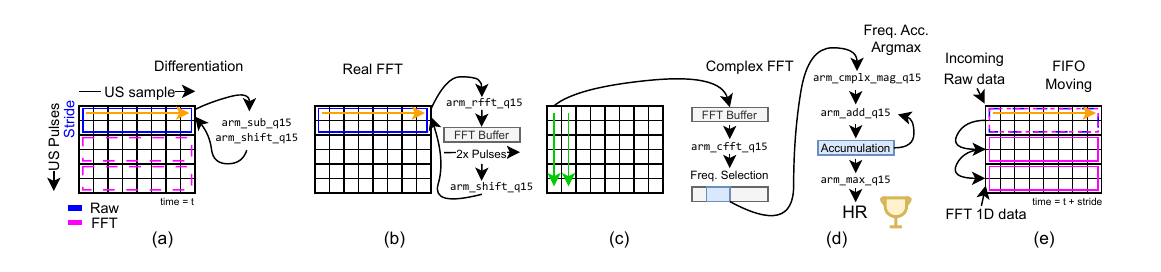}
    \caption{This figure illustrates the memory mapping process with the corresponding sequence of operations for the embedded algorithm. Panel (a) depicts the differentiation operation conducted along the \ac{us} sampling time axis. A \texttt{FFT Buffer} is allocated with double the memory requirement for reuse in the subsequent real \ac{fft}, shown in panel (b).  Panel (c) visualizes the computation of the complex \ac{fft} on the first half of the samples, leveraging the conjugate symmetry of the \ac{fft}. The magnitude of the \ac{fft} is then calculated and accumulated, followed by peak detection within the chosen frequency window, as shown in panel (d). Finally, panel (e) depicts the advancement of the main \texttt{Collection Buffer}.}
    \label{fig:memory_mapping}
\end{figure*}

\section{Embedded Implementation}
\label{sec:methods}
We present the architecture of our proposed wrist-worn energy-efficient wearable \ac{us} \ac{iot} in (Figure \ref{fig:schematic_operation}(a)) leveraging an ARM Cortex-M4 \ac{mcu} (STM32, ST Microelectronics, Switzerland) with mixed-signal capabilities. For \ac{us} signal excitation, this \ac{mcu} is paired with a \ac{soa} low-power pulser from ST Microelectronics. An analog front-end, including \ac{opamp} and filters, reduces the frequency bandwidth of the acquired \ac{us} signals. Finally, embedded \ac{dsp} enables real-time \ac{hr} detection directly on the wrist. The following subsections detail the system design, the \ac{dsp} algorithm, and the profiling of our proposed embedded architecture.

\subsection{System Level Architecture}
\label{subsec:system_description}

Figure \ref{fig:schematic_operation}(a) depicts our system designed for minimal component usage. It comprises three key parts: the pulser, the \ac{mcu}, and the envelope filter. The pulser selected is the \texttt{STHVUP32} (ST Microelectronics, Switzerland), designed and optimized for low-power operation. The pulser requires a combination of low-voltage and high-voltage power supplies. The low-voltage supplies include two at \qty{1.8}{\volt}, one at \qty{3.3}{\volt}, and one at \qty{-3.3}{\volt}. Additionally, two high-voltage supplies of \qty{\pm15}{\volt} drive the bridges connected to the transducer. On the analog domain, the pulser shows two outputs, \texttt{XDCR}, where the transducer is connected, and \texttt{LVOUT}, where the echo is relayed. In the digital domain, the pulser presents some logic signals as \texttt{TXON}, which signals the start of the pulse, and the Trigger (\texttt{TRG}) signal, which is used to trigger the start of a pulse. The \texttt{STHVUP32} can be configured via QSPI and must be clocked between \qty{10}{\mega\hertz} and \qty{100}{\mega\hertz}. To minimize the components of the clock tree, we are supplying the pulser clock from one of the \ac{pll} components embedded in the \ac{mcu}, using the high-frequency clock output (\texttt{MCO} pin) of the \ac{mcu}. The \ac{mcu} of choice is the \texttt{STM32L496}  (ST Microelectronics, Switzerland). It is a low-power \ac{mcu} based on an ARM Cortex M4 processor, an architecture widely used in Bluetooth low-energy modules. The \ac{mcu} features three on-board 5 Msps 12-bit \ac{adc}s, with power consumption as low as \qty{200}{\micro\ampere}/Msps. We use one of the on-board \ac{adc}s to sample the \ac{us} signal at 4 Msps for a total of 50 samples per echo. Moreover, \qty{1}{\mega\byte} of flash and \qty{320}{\kilo\byte} of RAM on the \ac{mcu} allow us to deploy complex algorithms on the edge.

\subsection{Bandwidth Reduction Circuit}
\label{subsec:envelope}
This work investigates how an envelope stage in the \ac{us} \ac{rx} circuit affects the accuracy of \ac{hr} detection. By reducing the sampling bandwidth required for high-frequency \ac{us} signals, the envelope stage facilitates the use of lower-power and lower-cost electronics. In particular, we use the integrated \ac{adc} of the \ac{mcu} to digitize the bandwidth-reduced data. Moreover, limiting the amount of acquired digital data minimizes the system's memory footprint and allows for the implementation of faster processing algorithms. This translates to reduced computational demands and less complex hardware requirements, ultimately enhancing the overall system efficiency. Thus, it can significantly contribute to embedding \ac{us} technology in wrist-based \ac{iot}s. 

We build on the circuit presented in Giordano et al.\cite{giordano_towards_2023} in which an envelope stage was used to reduce the bandwidth required for sampling \ac{us} signals. We have adapted the circuit proposed by the authors, adding a high-pass filter stage before the first amplification stage. The bottom half of Figure \ref{fig:schematic_operation}(a) shows a schematic of the circuit used in this work. The \texttt{ADA4807} (Analog Devices Inc., USA) \ac{opamp} has been chosen. It boasts rail-to-rail operation and an high bandwidth-product gain, making it the ideal candidate for the initial stages of the amplification circuit. Raw \ac{us} signals are fed into the high pass filter composed of Rhp and Chp, which removes any DC bias the signal might have. U1 and U2 are two \ac{opamp}s set in a non-inverting amplifier configuration, whose gain factors are defined by $1+\frac{R1}{Rf1}$ and $1+\frac{R2}{Rf2}$, respectively. The amplified signal is then fed into U3, a voltage follower that decouples the impedance network of the envelope detector from the feedback loop of U2. The envelope detector comprises the diode D1, which conducts only the positive part of the signal, and the capacitor C1, which gets charged and discharged by two resistors, R3 and R4. The resistor limits the capacitor's charging and discharging rate, allowing the capacitor's voltage to track the signal's envelope effectively. A low pass filter (Rlp, Clp) is added before the last signal amplification stage (the non-inverting \ac{opamp} U4).

\begin{figure*}[tbh]
    \centering
    \includegraphics[width=\linewidth]{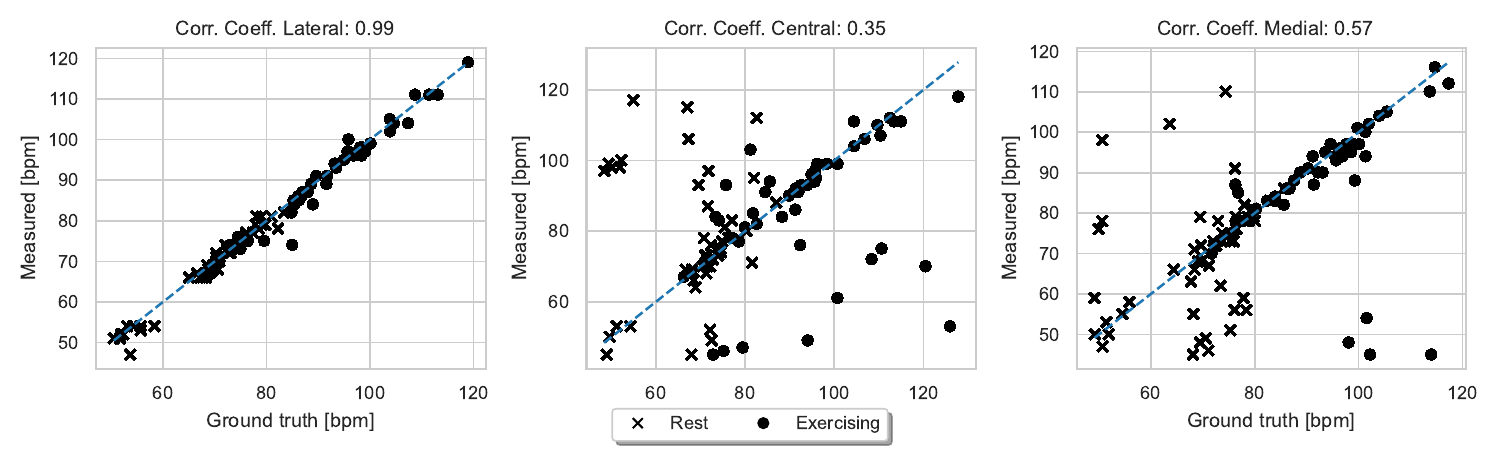}
    \caption{Correlation plots between the HR extracted from the \texttt{Polar H10} chest belt and the HR extracted from the enveloped \ac{us} signal at different positions on the wrist. The plots show the medial, central and lateral positions from left to right, respectively.}
    \label{fig:corr_plot}
\end{figure*}

\begin{figure*}[tbh]
    \centering
    \includegraphics[width=\linewidth]{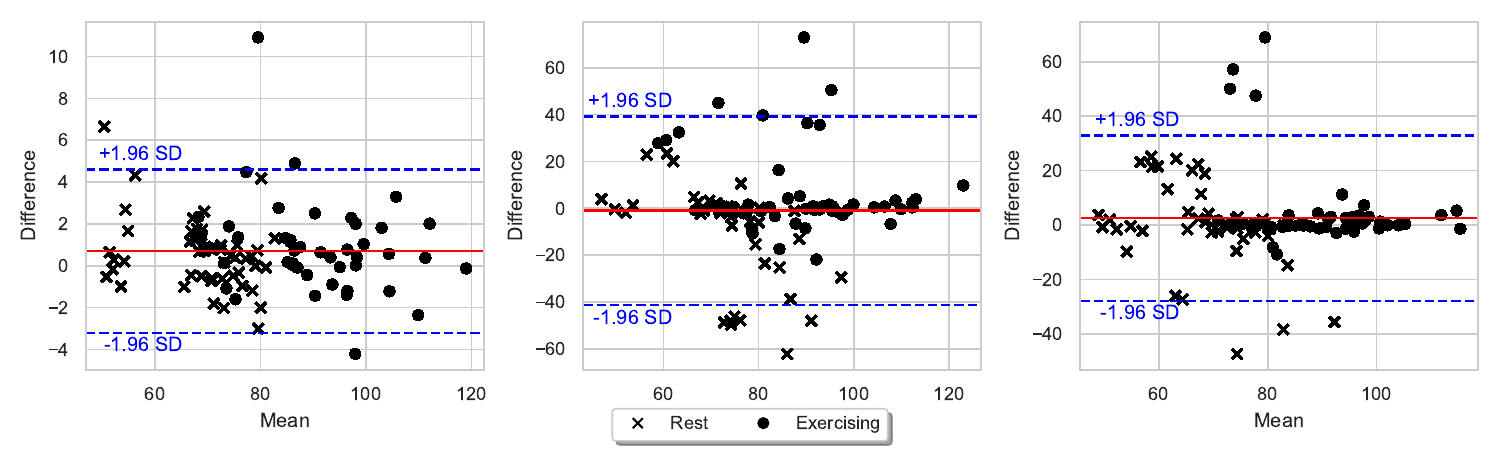}
    \caption{Bland-Altman plot between the HR extracted from the \texttt{Polar H10} chest belt and the HR extracted from the enveloped \ac{us} signal at different positions on the wrist. The plots show the medial, central and lateral positions from left to right, respectively.}
    \label{fig:bland_altman}
\end{figure*}

\subsection{System Operation}
\label{subsec:system_op}

Figure \ref{fig:schematic_operation}(b) shows a simplified software flowchart of the system. The operation begins at startup, initialling the \ac{mcu} peripherals, the pulsar's clock and setting its register via SPI. The system is then set in low-power Stop2 mode. Stop2 mode has been chosen for being the lowest low-power mode, allowing for full RAM retention, where the \ac{us} buffer is stored. The low power timer (LPTimer) wakes up the \ac{mcu} from the low power mode and is clocked at \qty{25}{Hertz}, effectively setting the \ac{us} pulse repetition frequency. When the system is woken up, the pulser's clock is re-enabled and a pulse is triggered via the timer interrupt. An ADC conversion is then initiated via the \texttt{TXON} pin, asserted by the pulser. \ac{dma} will handle the transfer between the ADC registers and memory buffer. Once the \ac{us} frame is saved, if enough samples have been collected to satisfy the set stride, the \acs{dsp} is run before the system is set back to low-power mode.

\subsection{On-board Signal Processing}
\label{subsec:onboard_signal_processing}

This section details the implementation of the signal processing algorithm described in Section~\ref{sec:preliminaries} on the \ac{mcu} utilizing optimized CMSIS-DSP kernels. Figure~\ref{fig:memory_mapping} illustrates the data flow and corresponding function calls associated with each \ac{dsp} step. Careful consideration was given to buffer allocation due to the performance implications of data movement and the limited memory available on the MCU, which restricts data duplication. Consequently, three buffers are employed and stored as RAM arrays on the MCU:
\begin{enumerate}
    \item \textbf{Main} \texttt{Collection Buffer:} This buffer has the dimensions of $\text{\#pulses}\cdot\text{\#samples}$ and stores the recorded \ac{us} \ac{rx} data.
    \item \textbf{Temporary} \texttt{FFT Buffer:} This buffer has a dimension of $2\cdot\text{\#pulses}$ and serves as temporary storage for the \ac{fft} calculation.
    \item \textbf{\ac{hr}} \texttt{Accumulation Buffer:} This buffer has a dimension of $1\cdot\text{\#Freqs}$, where $\text{\#Freqs}$ represents the number of discrete filtered \ac{hr} frequencies. 
\end{enumerate}

\begin{table}[htb]
    \centering
    \begin{center}
    \begin{tabular}{cccc}
        \toprule
        Wrist Pos. & Corr. Coeff. & Mean Err.[bpm] & Std. Dev.[bpm] \\
        \midrule
        Lateral & 0.99 & 0.69 & 1.99 \\
        Central & 0.35 & -1.04 & 20.54 \\
        Medial & 0.57 & 2.50 & 15.55 \\
        \bottomrule
    \end{tabular}
    \end{center}
    \caption{Statistics on \ac{hr} accuracy for different wrist positions from all subjects in the dataset}
    \label{tab:hr_accuracy}
\end{table}

The main \texttt{Collection Buffer} functions as a \ac{fifo} buffer. It stores incoming \ac{us} \ac{rx} signals, overwriting the oldest data points that remain from the previous \ac{dsp} instance at time \qty{}{t}+\qty{}{stride} (Figure \ref{fig:memory_mapping}(e)). At a new \ac{dsp} instance triggered at a frequency of $1/stride$, the function calls are executed on the newly acquired \ac{adc} values within the stride~\footnote{e.g. corresponding to 50 data rows at \ac{prt} of \qty{25}{\hertz} and a chosen stride length of \qty{2}{\second}}, resulting in a new \ac{hr} reading. The on-board processing is achieved through five consecutive steps, as depicted in Figure \ref{fig:memory_mapping}(a)-(e):

\begin{itemize}
    \item \textbf{Differentiation}: the initial processing begins with $arm\_shift\_q15$ and $arm\_sub\_q15$ functions. The $arm\_shift\_q15$ operation expands the acquired 12-bit \ac{adc} data to utilize the full 16-bit width of the buffer. This operation modifies data directly within the main collection buffer, using it as both source and destination, as shown in Figure \ref{fig:memory_mapping}(a).
    \item \textbf{Real \ac{fft}}: Following differentiation, the first \ac{fft}, termed \texttt{Real FFT} due to its operation on real \ac{adc} values, is computed using the $arm\_rfft\_q15$ function. This function necessitates the storage of intermediate results in a temporary buffer, designated \texttt{FFT Buffer}, as it modifies the input data. After the computation, the buffer contents are copied back to the main collection buffer. Although both real and imaginary components of the \ac{fft} are computed and stored in the temporary \ac{fft} buffer, only half the frequency spectrum needs to be retained and copied back into the main collection buffer due to the inherent conjugate symmetry. Therefore, the memory requirements do not change.
    
    Moreover, it is noteworthy that the size of the \texttt{FFT Buffer} is $2\cdot\text{\#pulses}$, exceeding the minimum memory requirement for the first FFT  ($2\cdot\text{\#samples}$). This larger size is chosen for efficient reuse of the same buffer in the subsequent processing step. 
    \item \textbf{Complex \ac{fft}}: the second \ac{fft} follows and must be computed along the columns. Since the collection buffer is stored in a contiguous space in RAM, columns' matrixes must first be extracted and saved in a buffer, for which we re-use the \texttt{FFT Buffer}. The complex \ac{fft} is computed in place on the temporary buffer with the function $arm\_cfft\_q15$. The output of this step is graphed in Figure \ref{fig:memory_mapping}(c).
    \item \textbf{Frequency Selection}: Following the second \ac{fft}, the signal information now resides in the frequency domain of the \ac{hr} signal. To improve computational efficiency, we restrict the analysis to a predefined band of interest (as detailed in Section~\ref{sec:preliminaries}). This frequency selection is performed in-place using pointer arithmetic (Figure \ref{fig:memory_mapping}(c)).
    \item \textbf{Frequency Accumulation}: on the same \texttt{FFT Buffer}, the magnitude is then computed with the function $arm\_cmplx\_mag\_q15$. The results are summed on a second temporary buffer, the \texttt{Accumulation Buffer}, with the function $arm\_add\_q15$, which holds the accumulated frequencies\footnote{As the accumulation proceeds in a column-wise fashion along \ac{dsp} events, the accuracy of \ac{hr} extraction progressively improves as the buffer fills to capacity.}. These steps are presented in Figure \ref{fig:memory_mapping}(d).
    \item \textbf{Peak finding}: as a final step, the function $arm\_max\_q15$ finds the peak in the signal in the useful frequencies, which is then converted into \ac{hr}, as depicted in Figure \ref{fig:memory_mapping}(d).
    \item \textbf{FIFO advancement}: lastly, the \ac{fifo} advances by one stride value, moving the already computed Real FFT values and overwriting the oldest ones. It is now possible to acquire new data on the top of the \texttt{Collection Buffer}.
\end{itemize}

The embedded implementation of \ac{hr} detection necessitated further refinement of two key algorithm parameters: the window size and stride. The window size is defined as the number of \ac{us} frames stacked in one M-Mode representation. This parameter directly affects the memory consumption, the latency, and therefore, the power consumption of the algorithm, and the accuracy of the \ac{hr} extraction. In contrast, the window stride defines the interval between running the \ac{dsp} algorithm. This determines how frequently a new \ac{hr} measurement is provided. Unlike the window size, the stride primarily affects power consumption. The \ac{mcu} needs to wake up and perform the \ac{dsp} computation at each stride interval. Hence, stride does not impact memory and latency since the same amount of data always fills the window.

\subsection{Power Profiling And Performance Evaluation}
\label{subsec:power_profiling}

The system's power consumption was characterized using the \texttt{Nordic Power Profiler Kit II} (Nordic Semiconductor, Norway) \cite{nordicsemiPowerProfiler}. A granular approach was used, profiling each element individually: the \ac{us} pulser, the analog envelope circuit, and the \ac{mcu}. Each power domain was investigated separately for the \ac{us} pulser, and the individual power intakes were summed to obtain the overall pulser power consumption. For the voltage pairs \qty{\pm3.3}{\volt} and \qty{\pm15}{\volt} in the pulser, we assumed symmetrical power draw, profiled the positive voltage rails and added them in duplicate. Similarly, the \ac{opamp} within the envelope stage were profiled by monitoring their \qty{\pm5}{\volt} inputs. Concerning the \ac{mcu}, power consumption was characterized through Vcc, which was set to \qty{1.8}{\volt}.

To assess the performance of our embedded algorithm relative to reference \ac{ecg} signals, we computed the mean errors and standard deviations of \ac{hr} extraction upon achieving the full capacity of the main \texttt{Collection Buffer}.

\section{Results and Discussion}
\label{sec:results}

This section presents the findings of our experiments on extracting \ac{hr} from wrist-based \ac{us} data. We evaluated the effectiveness of different transducer positions on the wrist by analyzing the extracted \ac{hr} from the enveloped \ac{us} data. Subsequently, we analyzed the results from the embedded implementation, focusing on achieving an optimal trade-off between accuracy, latency, and power consumption. Finally, we provide a detailed performance breakdown of the algorithm on the embedded system, including power consumption, memory usage, and latency.

\begin{figure*}[thb]
    \centering
    \begin{minipage}{0.23\textwidth}
        \centering
        \includegraphics[width=\linewidth]{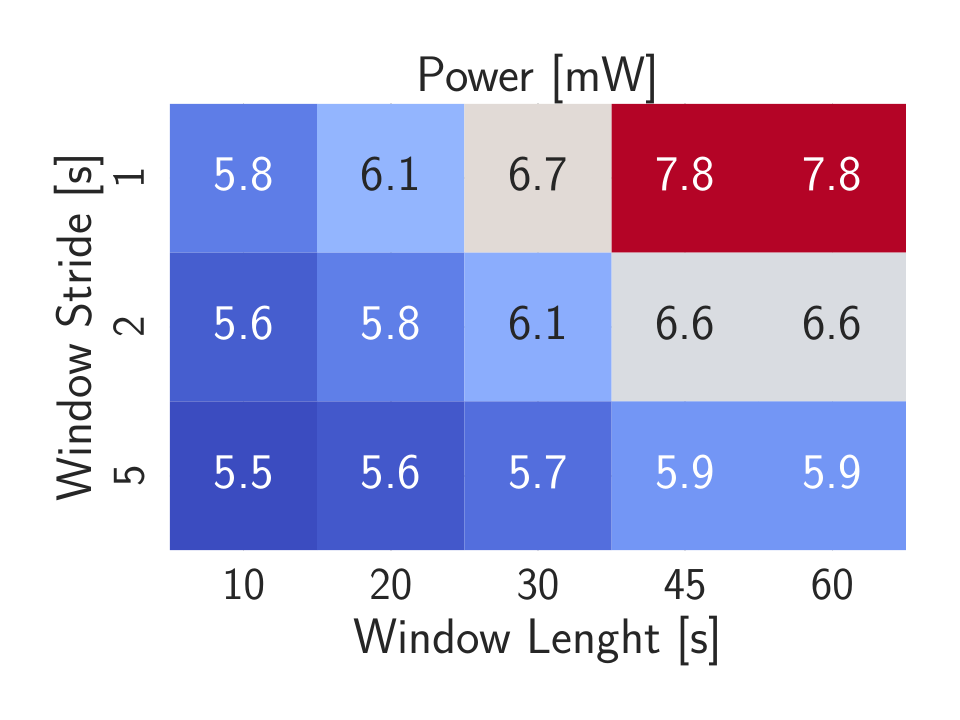}
        \subcaption{}
    \end{minipage}
    \hfill
    \begin{minipage}{0.23\textwidth}
        \centering
        \includegraphics[width=\linewidth]{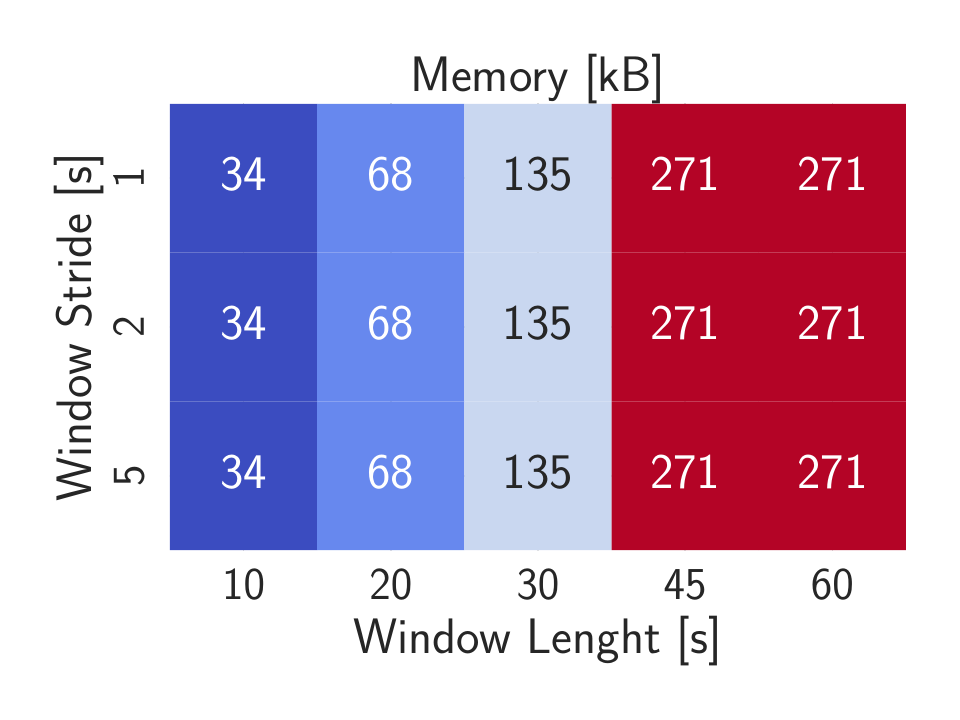}
        \subcaption{}
    \end{minipage}
    \hfill
    \begin{minipage}{0.23\textwidth}
        \centering
        \includegraphics[width=\linewidth]{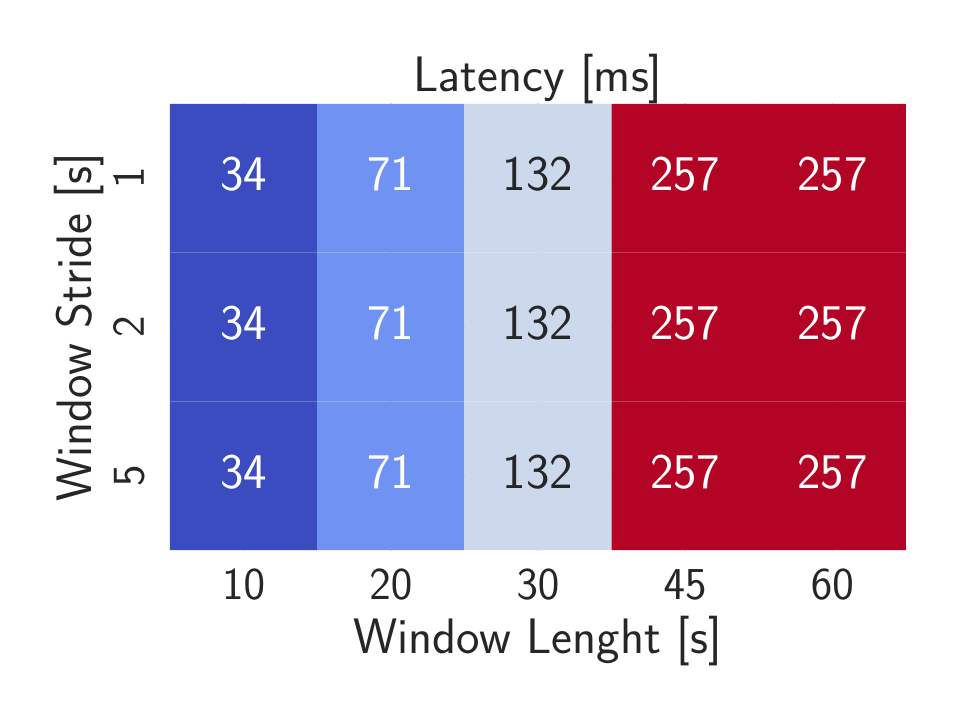}
        \subcaption{}
    \end{minipage}
    \hfill
    \begin{minipage}{0.23\textwidth}
        \centering
        \includegraphics[width=\linewidth]{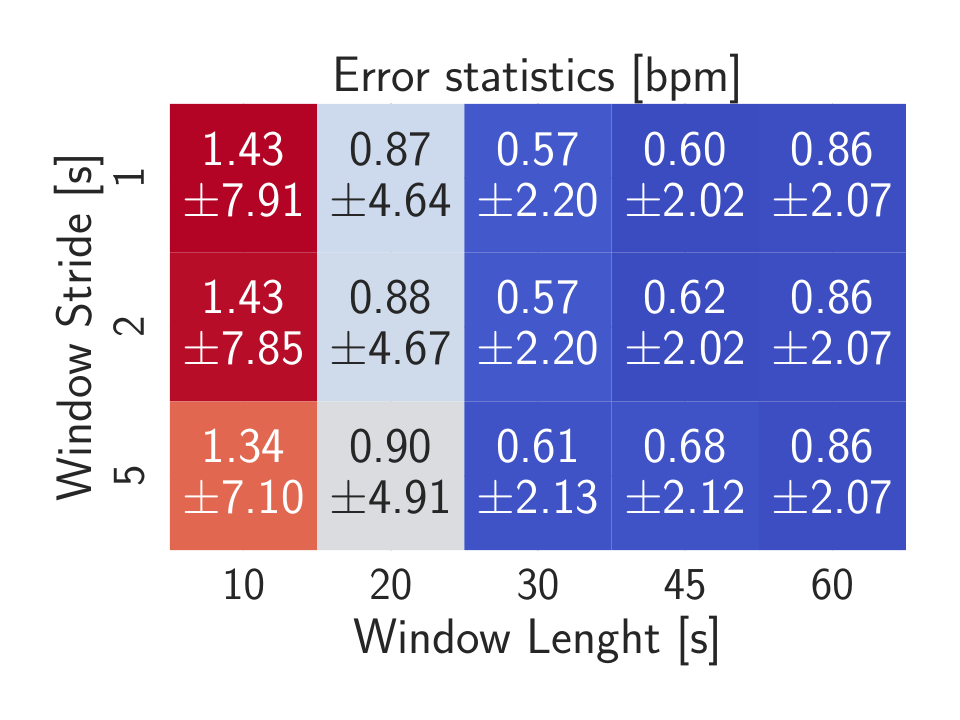}
        \subcaption{}
    \end{minipage}
    \caption{Analysis of the embedded algorithm's key performance metrics varying window length and stride. (a) shows the system's total power consumption with a pulse repetition frequency of \qty{25}{\hertz}. (b) shows the memory footprint of the algorithm. (c) shows the latency of the \ac{dsp} execution. (d) shows the two algorithm parameters' impact on \ac{hr} accuracy profiled at buffer capacity.}
    \label{fig:heatmaps}
\end{figure*}

\begin{figure*}[thb]
    \begin{minipage}{0.28\textwidth}
            \centering
            \vspace{-0.25cm}
        \includegraphics[width=\textwidth]{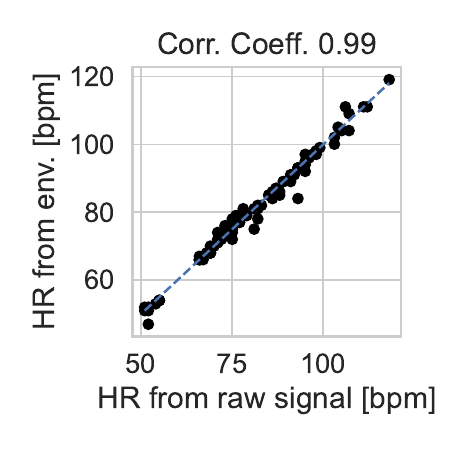}
                \vspace{-0.25cm}
                \caption{Correlation plot between \ac{hr} extracted from the raw and the enveloped \ac{us} signal.}
                \label{fig:env_lvout_corr}
    \end{minipage}
    \hfill
    \begin{minipage}{0.7\textwidth}
            \centering
            \includegraphics[width=\textwidth,trim=0cm 0cm 0cm 0cm,clip]{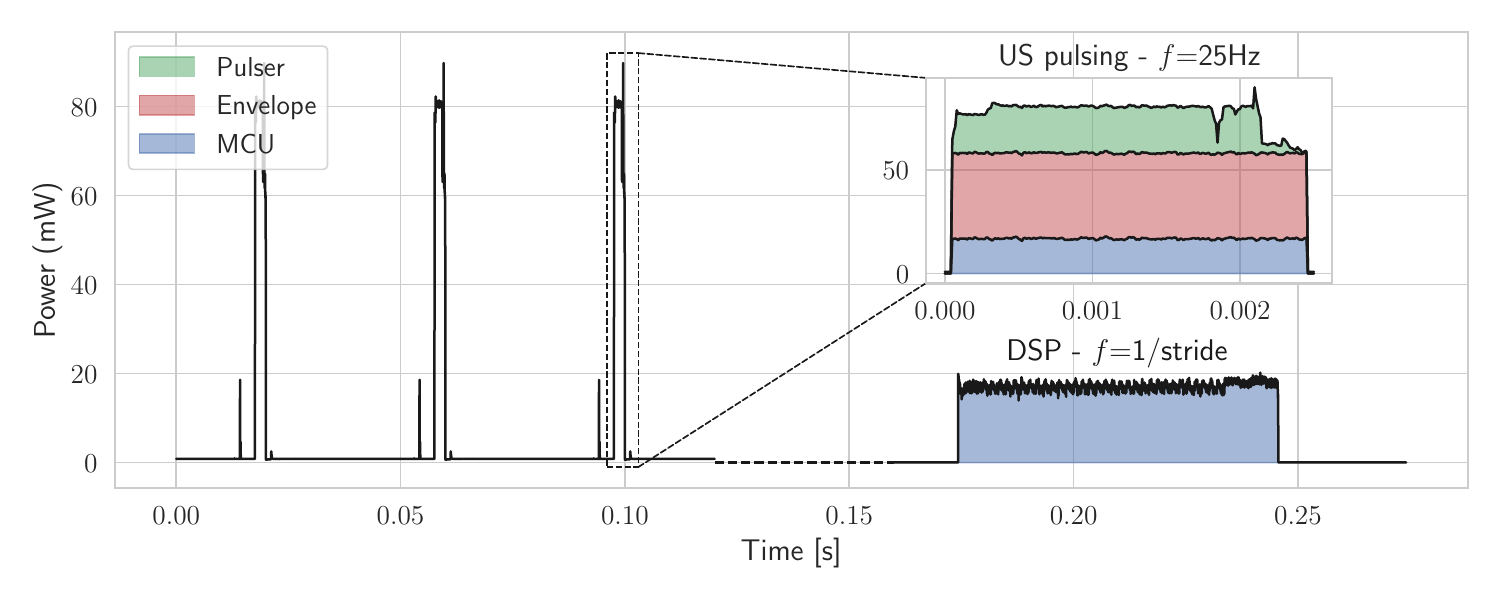}
            \caption{Power profiling of the operation. The spikes at the left indicate the periodic wake-up of the system and the ultrasound pulse. In the center, a zoomed-in version details the contribution to the power consumption of the different subsystems. The DSP is depicted on the right side of the plot, and the algorithm's stride sets its period.}
            \label{fig:power_profile}
    \end{minipage}
\end{figure*}

\subsection{HR Extraction}
As a first step, we investigated the optimal wrist position for \ac{hr} data collection. Therefore, we extracted the \ac{hr} from the enveloped \ac{us} signals, following the signal processing steps introduced in Section \ref{sec:preliminaries}, captured from the three different positions on the wrist as detailed in Section \ref{sec:dataset}. We compared the computed \ac{hr} with the reference \ac{hr} signal obtained from the \ac{ecg} chest belt. The correlation and Bland-Altman plots are shown in Figure \ref{fig:corr_plot} and Figure \ref{fig:bland_altman}, respectively. The correlation coefficient, mean error, and standard deviation from the ECG readings are reported in Table \ref{tab:hr_accuracy}.

Bland-Altman plots in Figure \ref{fig:bland_altman} show that our method employing enveloped \ac{us} has no substantial bias in the lateral (\qty{0.69}{}$\pm$\qty{1.99}{bpm}), central (\qty{-1.04}{}$\pm$\qty{20.54}{bpm}) and medial (\qty{2.50}{}$\pm$\qty{15.55}{bpm}) wrist positions commpared to \ac{ecg}. Furthermore, a statistically significant positive correlation was observed between \ac{ecg} and enveloped \ac{us} derived \ac{hr} for all placements: lateral (r(92)=0.99, p$<$0.001), central (r(98)=0.35, p$<$0.001) and medial (r(98)=0.57, p$<$0.001). Notably, the lateral placement near the radial artery displayed the strongest correlation and the lowest mean difference in the Bland-Altman plot. This aligns with the high \ac{snr} typically observed in this area due to the superficial positioning of the radial artery, mirroring the advantage of manual palpation for HR detection. These findings demonstrate the potential of \ac{us} for accurate \ac{hr} extraction, achieving comparable performance to established wearable technology \cite{comaprison_ppg}.

Analysis of Figures~\ref{fig:corr_plot} and~\ref{fig:bland_altman} for the central and medial position suggests a potential association between lower measured \ac{hr} during resting state and inaccurate \ac{hr} reconstruction using \ac{us}. This observation could be attributed to a decreased \ac{snr} in the \ac{us} signal at lower \ac{hr}s due to weaker cardiovascular function. Furthermore, the respiration rate harmonics are known to lie in the \qty{40}{}-\qty{60}{bpm}, with resting respiratory rates in healthy individuals typically between \qty{10}-\qty{15}{bpm}\cite{kral2023slower}. Consequently, in wrist positions where blood vessels are less prominent, respiration harmonics might overpower the \ac{hr} signal, potentially explaining the observed cluster of inaccurate measurements around \qty{60}{}-\qty{90}{bpm} in the correlation plots (Figure \ref{fig:corr_plot}).

After establishing the feasibility of \ac{hr} extraction using \ac{us} and identifying the optimal measurement location, the next step was to evaluate the impact of signal bandwidth reduction through envelope detection on \ac{hr} accuracy. Figure \ref{fig:env_lvout_corr} demonstrates a strong correlation (r(92)=0.99, p$<$0.001) between \ac{hr} extracted from the enveloped and raw signals. This confirms the effectiveness of the enveloping technique in reducing signal bandwidth without compromising \ac{hr} accuracy. Moreover, using lower-bandwidth signals allows for reducing the memory footprint and computational complexity of the algorithm, as well as the power consumption. This also relaxes the requirements for \ac{adc} performance while still adhering to Nyquist's theorem. Consequently, \ac{us} signals can be acquired with inexpensive and low-power \ac{adc}s, including on-board \ac{adc}s commonly found on \ac{cots} \ac{mcu}s, as demonstrated in this work.

\begin{table}[htb]
    \centering
    \begin{center}
    \begin{tabular}{cccc}
        \toprule
        Metric & Float$^\dagger$ & Fixed q1.15$^\dagger$ & Improvement$^\ddagger$\\
        \midrule
        Latency [ms] & 90 & 71 & 21\% \\
        Memory [kB] & 136 & 68 & 50\% \\\
        Energy [mJ] & 1.71 & 1.21 & 29\% \\
        \bottomrule
    \end{tabular}
    \end{center}
    \begin{tablenotes}
        \item $^\dagger$ Computed for a window length of \qty{20}{\second} and a stride length of \qty{2}{\second}
        \item $\ddagger$ Relative difference between floating- and fixed-point implementation.
    \end{tablenotes}
    \caption{Metrics of the floating-point and fixed-point implementations of our embedded algorithm. The characteristics were profiled for the chosen window length of \qty{20}{\second} and stride length of \qty{2}{\second}.}
    \label{tab:float_fixed}
\end{table}

\subsection{On-board Processing}
The system's average power consumption, memory usage, and latency have been measured for different window sizes and stride values. The results are shown in Figure \ref{fig:heatmaps}(a), Figure \ref{fig:heatmaps}(b), and Figure \ref{fig:heatmaps}(c), respectively. Moreover, the signal processing has been run for different window sizes and stride values, yielding different accuracies, as detailed in Figure \ref{fig:heatmaps}(d).

Porting the algorithm in an embedded \ac{mcu} requires fine-tuning its parameters to achieve the optimal balance between accuracy, latency, and power consumption. Figure~\ref{fig:heatmaps} shows that, as expected, longer windows enhance accuracy but come at the cost of increased memory footprint, latency, and, ultimately, power consumption. Window stride, on the other hand, does not impact memory consumption and latency. This is because the size of the buffers storing the signal and the underlying \ac{dsp} algorithm remains the same. However, window stride directly influences the execution frequency of the \ac{dsp} algorithm, affecting power consumption. Moreover, window stride slightly affects accuracy because the algorithm is run on very similar data (the stride accounts for the new data).

We determined the best trade-off between the algorithm's accuracy and the system's low-power requirements with a window length of \qty{20}{\second} and a window stride of \qty{2}{\second}. This setting allows to achieve a mean error of \qty{0.88}{bpm} and a standard deviation of \qty{4.67}{bpm}, with a latency of \qty{71}{\milli\second}, a total memory consumption of \qty{68}{\kilo\byte} and an average power consumption of \qty{5.8}{\milli\watt}.

Our investigations included also a comparative analysis of both floating-point and fixed-point implementations of the embedded algorithm. Table \ref{tab:float_fixed} summarizes the results. The fixed-point implementation exhibited significant performance improvements in both latency and energy consumption. Compared to the floating-point implementation, latency was reduced by 21.0\%, and energy consumption by 29.0\%. Furthermore, the fixed-point implementation offers substantial memory efficiency, requiring only 50\% of the memory footprint due to its reduced data size (\qty{2}{\byte} per sample for \texttt{q1.15} format vs. \qty{4}{\byte} per sample for \texttt{fp32} format).

\subsection{Power Consumption}
Window size and window stride had to be set before measuring the system's power consumption. The configuration providing the best balance between memory utilisation, latency and \ac{hr} accuracy in terms of power consumption was determined with a window size of \qty{20}{\second} and a step size of \qty{2}{\second}. Figure~\ref{fig:power_profile} shows the power profile over time for these settings. The left part of the Figure depicts the system's power consumption during the pulse generation, the echo enveloping, and the acquisition of the \ac{us} signal, singularly detailed in the zoomed-in plot. This operation is run periodically every \qty{40}{\milli\second}, yielding a pulse repetition frequency (\ac{us} frames per second) of \qty{25}{\hertz}. Once per window stride, the \ac{mcu} is woken up, and the \ac{dsp} is run. The plot on the right side shows the power consumption of the \ac{mcu} during the extraction of the \ac{hr}.

Having fixed the algorithm's parameters, we thoroughly power-profiled the system, as shown in Figure \ref{fig:power_profile}. The left half of the plot depicts the power consumption during the \ac{us} pulse and signal acquisition phase. The system is duty-cycled to achieve the lowest power draw possible, as detailed in Section \ref{subsec:system_description}. A breakdown of the power consumption during the acquisition phase is shown in the center part of the plot: the envelope stage consumed 53\% of power, and the MCU and pulser 25\% and 22\%, respectively. Our analysis identified the envelope stage as the primary contributor to system power consumption. This is attributed to the selection of an \ac{opamp} with a higher bandwidth-gain product than strictly necessary for the application, resulting in increased power draw. Nevertheless, the envelope stage was indispensable for this study. Without its bandwidth reduction capabilities, the \ac{mcu} would be unable to sample the signal, thus requiring a significantly more power-hungry \ac{adc} and a processor to handle the increased data volume. To explore potential power reductions, simulations were conducted using components with lower performance and power characteristics. For instance, replacing the last two stages of the envelope circuit (U3 and U4 in Figure \ref{fig:schematic_operation}(a)) with the \texttt{LTC6162} results in average power consumption of \qty{24.8}{\milli\watt} in the ON state, compared to the measured \qty{39.6}{\milli\watt} in the current setup.

The final third of the plot in Figure \ref{fig:power_profile} depicts power consumption during the signal processing stage. The \ac{dsp} benefits from the efficient implementation of mathematical functions provided by the CMSIS-DSP libraries, and the window stride determines its execution frequency. Notably, the \ac{mcu} maintains a constant power draw of around \qty{20}{\milli\watt} throughout the entire \ac{hr} extraction process, which takes about \qty{200}{\milli\second}.

Furthermore, a commonly used commercial \ac{ppg} \ac{ic}, the Analog Devices \texttt{AX30102} (Analog Devices, USA), reportedly consumes up to \qty{1.2}{\milli\watt} at its lowest sampling rate of \qty{50}{sps}. However, this value is limited to the sensing process itself, and additional power is required for sensor data retrieval and processing to extract \ac{hr}. In contrast, our proposed solution achieves an average power consumption of only \qty{5.8}{\milli\watt} for the entire application. Moreover, we estimate that our system's battery lifetime, operating continuously under the aforementioned parameters, is seven days on a single charge. This estimation is based on a \qty{309}{\milli\ampere\hour} battery with a nominal voltage of \qty{3.85}{\volt}, the same as the one used in the Apple Watch Series 7 \cite{AppleWat87:online}.

\section{Limitations and Future Work}
\label{sec:limitations}

The dataset for this work was collected with participants in a static, seated position, while real-world scenarios might involve wrist and hand movements that can introduce artefacts into the sensor data. However, the form factor of a wristwatch, combined with the necessity of a watchband, inherently creates a close coupling between the transducer and the skin. This natural coupling can help mitigate the effects of motion artefacts. Additionally, in the collection of our dataset, we have already incorporated higher \ac{hr} (after light physical activity) to account for exercise conditions. Moreover, our data collection utilized a \ac{us} gel pad to achieve acoustic coupling between the wrist and the transducer. This approach may pose limitations for real-world applications. However, soft elastomeric materials such as \ac{pdms}, rubber or silicone are commercially available and are often used in watch bands \cite{swatch_replacement_nodate}. These materials exhibit suitable acoustic properties, suggesting their potential to facilitate the seamless integration of transducers and replicate the efficient coupling achieved by the gel pad. Furthermore, it should be mentioned that this study does not consider complete system integration and miniaturization at the \ac{ic} level, which would require considerable technical effort and is beyond the investigatory scope of this work.

Future research on wearable \ac{us} technology on the wrist holds promise for non-invasive measurement of vital parameters that are currently difficult to monitor continuously, such as blood pressure \cite{8478346} and blood glucose \cite{park2020ultrasound}. In addition, wrist-worn \ac{us} has the potential to be exploited for next-generation human-machine interfaces due to its ability to penetrate the skin and observe muscle and tendon movements. This could lead to applications in smartwatches and smart bracelets, enabling one-handed control. Beyond that, \ac{us} technology could be integrated into augmented and extended reality applications, offering a more power-efficient and less computationally demanding alternative to camera-based input methods.
\section{Conclusion}
\label{sec:conclusion}

This work investigated the feasibility and potential of using low-power \ac{us} technology for measuring \ac{hr} on the wrist in an ultra-low-power fashion. Our study achieved high accuracy by extracting \ac{hr} data from 10 participants using a single A-mode US channel. Compared to \ac{ecg}-based \ac{hr} measurements, the results showed a strong correlation (r(92)=0.99, p$<$0.001) and a mean error of \qty{0.69}{}$\pm$\qty{1.99}{bpm}. Additionally, the study identified the area around the radial artery on the wrist as the optimal position for the transducer to achieve the highest \ac{snr}.

To incorporate \ac{us} technology into wearable devices, we demonstrated that enveloping the high-frequency \ac{us} signal can lead to a significant ($>5\times$) bandwidth reduction without affecting the accuracy of the \ac{hr} extraction algorithm. This technique enables the use of low-power ARM Cortex-M \ac{mcu}s for signal acquisition. This significantly reduces the cost and complexity of acquisition front-ends while still providing \ac{hr} measurement accuracies that closely match the \ac{hr} extracted from raw \ac{us} signals (correlation coefficient, r(92)=0.99, p$<$0.001). Moreover, the signal processing required to extract \ac{hr} from A-mode \ac{us} signals has been ported in fixed-point arithmetic requiring only 68 kB of RAM and \qty{71}{\milli\second} for execution on a Cortex-M4 based \ac{mcu}. Furthermore, the system has been fully power-profiled, yielding an average power consumption of \qty{5.8}{\milli\watt} and translating to an estimated battery life of $>7$ days for continuous monitoring when equipped with a standard smartwatch battery.

\section*{Acknowledgment}
 The authors acknowledge support from the Swiss National Science Foundation (Project PEDESITE) under grant agreement 193813 and from the ETH Research Grant ETH-C-01-21-2 (Project ListenToLight). Furthermore, the authors would like to acknowledge Andreas Habersack for discussions and considerations in the design of the data collection setup as well as Martin Tanner for his help in the circuit profiling. Moreover, we want to thank ST Microelectronics for having supplied the hardware and sharing their expertise in \ac{us} pulser design.

\bibliographystyle{IEEEtran}
\bibliography{ultrasound_hr}

\begin{thebibliography}{10}
\providecommand{\url}[1]{#1}
\csname url@samestyle\endcsname
\providecommand{\newblock}{\relax}
\providecommand{\bibinfo}[2]{#2}
\providecommand{\BIBentrySTDinterwordspacing}{\spaceskip=0pt\relax}
\providecommand{\BIBentryALTinterwordstretchfactor}{4}
\providecommand{\BIBentryALTinterwordspacing}{\spaceskip=\fontdimen2\font plus
\BIBentryALTinterwordstretchfactor\fontdimen3\font minus
  \fontdimen4\font\relax}
\providecommand{\BIBforeignlanguage}[2]{{%
\expandafter\ifx\csname l@#1\endcsname\relax
\typeout{** WARNING: IEEEtran.bst: No hyphenation pattern has been}%
\typeout{** loaded for the language `#1'. Using the pattern for}%
\typeout{** the default language instead.}%
\else
\language=\csname l@#1\endcsname
\fi
#2}}
\providecommand{\BIBdecl}{\relax}
\BIBdecl

\bibitem{ensminger_ultrasonics_2024}
\BIBentryALTinterwordspacing
D.~Ensminger and L.~J. Bond, \emph{Ultrasonics: {Fundamentals}, {Technologies},
  and {Applications}}, 4th~ed.\hskip 1em plus 0.5em minus 0.4em\relax Boca
  Raton: CRC Press, Feb. 2024. [Online]. Available:
  \url{https://www.routledge.com/Ultrasonics-Fundamentals-Technologies-and-Applications/Ensminger-Bond/p/book/9780367252816}
\BIBentrySTDinterwordspacing

\bibitem{9691276}
L.~Jia, L.~Shi, C.~Liu, J.~Xu, Y.~Gao, C.~Sun, S.~Liu, and G.~Wu,
  ``Piezoelectric micromachined ultrasonic transducer array-based electronic
  stethoscope for internet of medical things,'' \emph{IEEE Internet of Things
  Journal}, vol.~9, no.~12, pp. 9766--9774, 2022.

\bibitem{silva2023}
\BIBentryALTinterwordspacing
M.~Inês~Silva, E.~Malitckii, T.~G. Santos, and P.~Vilaça, ``Review of
  conventional and advanced non-destructive testing techniques for detection
  and characterization of small-scale defects,'' \emph{Progress in Materials
  Science}, vol. 138, p. 101155, Sep. 2023. [Online]. Available:
  \url{https://www.sciencedirect.com/science/article/pii/S0079642523000877}
\BIBentrySTDinterwordspacing

\bibitem{fitzpatrick_airborne_2020}
\BIBentryALTinterwordspacing
A.~Fitzpatrick, A.~Singhvi, and A.~Arbabian, ``An {Airborne} {Sonar} {System}
  for {Underwater} {Remote} {Sensing} and {Imaging},'' \emph{IEEE Access},
  vol.~8, pp. 189\,945--189\,959, 2020, conference Name: IEEE Access. [Online].
  Available: \url{https://ieeexplore.ieee.org/abstract/document/9228880}
\BIBentrySTDinterwordspacing

\bibitem{schulthess_passive_2024}
\BIBentryALTinterwordspacing
L.~Schulthess, P.~Mayer, L.~Benini, and M.~Magno, ``A {Passive} and
  {Asynchronous} {Wake}-up {Receiver} for {Acoustic} {Underwater}
  {Communication},'' May 2024, arXiv:2405.18000 [cs, eess]. [Online].
  Available: \url{http://arxiv.org/abs/2405.18000}
\BIBentrySTDinterwordspacing

\bibitem{szabo_diagnostic_2014}
\BIBentryALTinterwordspacing
T.~L. Szabo, \emph{Diagnostic {Ultrasound} {Imaging}: {Inside} {Out} ({Second}
  {Edition})}.\hskip 1em plus 0.5em minus 0.4em\relax Boston: Academic Press,
  Jan. 2014. [Online]. Available:
  \url{https://www.sciencedirect.com/science/article/pii/B9780123964878000185}
\BIBentrySTDinterwordspacing

\bibitem{jin_estimation_2024}
\BIBentryALTinterwordspacing
Y.~Jin, J.~T. Alvarez, E.~L. Suitor, K.~Swaminathan, A.~Chin, U.~S. Civici,
  R.~W. Nuckols, R.~D. Howe, and C.~J. Walsh,
  ``\BIBforeignlanguage{en}{Estimation of joint torque in dynamic activities
  using wearable {A}-mode ultrasound},'' \emph{\BIBforeignlanguage{en}{Nature
  Communications}}, vol.~15, no.~1, p. 5756, Jul. 2024, publisher: Nature
  Publishing Group. [Online]. Available:
  \url{https://www.nature.com/articles/s41467-024-50038-0}
\BIBentrySTDinterwordspacing

\bibitem{mohammadzadeh_asl_beamforming_2024}
\BIBentryALTinterwordspacing
B.~Mohammadzadeh~Asl and R.~Paridar, \emph{\BIBforeignlanguage{en}{Beamforming
  in {Medical} {Ultrasound} {Imaging}}}, ser. Springer {Tracts} in {Electrical}
  and {Electronics} {Engineering}.\hskip 1em plus 0.5em minus 0.4em\relax
  Singapore: Springer Nature, 2024. [Online]. Available:
  \url{https://link.springer.com/10.1007/978-981-99-7528-0}
\BIBentrySTDinterwordspacing

\bibitem{daniels_practical_2016}
\BIBentryALTinterwordspacing
J.~M. Daniels and R.~A. Hoppmann, Eds., \emph{\BIBforeignlanguage{en}{Practical
  {Point}-of-{Care} {Medical} {Ultrasound}}}.\hskip 1em plus 0.5em minus
  0.4em\relax Cham: Springer International Publishing, 2016. [Online].
  Available: \url{https://link.springer.com/10.1007/978-3-319-22638-5}
\BIBentrySTDinterwordspacing

\bibitem{wang2022}
\BIBentryALTinterwordspacing
C.~Wang, X.~Chen, L.~Wang, M.~Makihata, H.-C. Liu, T.~Zhou, and X.~Zhao,
  ``Bioadhesive ultrasound for long-term continuous imaging of diverse
  organs,'' \emph{Science}, vol. 377, no. 6605, pp. 517--523, Jul. 2022,
  publisher: American Association for the Advancement of Science. [Online].
  Available: \url{https://www.science.org/doi/full/10.1126/science.abo2542}
\BIBentrySTDinterwordspacing

\bibitem{hu2023}
\BIBentryALTinterwordspacing
H.~Hu, Y.~Ma, X.~Gao, D.~Song, M.~Li, H.~Huang, X.~Qian, R.~Wu, K.~Shi,
  H.~Ding, M.~Lin, X.~Chen, W.~Zhao, B.~Qi, S.~Zhou, R.~Chen, Y.~Gu, Y.~Chen,
  Y.~Lei, C.~Wang, C.~Wang, Y.~Tong, H.~Cui, A.~Abdal, Y.~Zhu, X.~Tian,
  Z.~Chen, C.~Lu, X.~Yang, J.~Mu, Z.~Lou, M.~Eghtedari, Q.~Zhou, A.~Oberai, and
  S.~Xu, ``\BIBforeignlanguage{en}{Stretchable ultrasonic arrays for the
  three-dimensional mapping of the modulus of deep tissue},''
  \emph{\BIBforeignlanguage{en}{Nature Biomedical Engineering}}, vol.~7,
  no.~10, pp. 1321--1334, Oct. 2023, publisher: Nature Publishing Group.
  [Online]. Available: \url{https://www.nature.com/articles/s41551-023-01038-w}
\BIBentrySTDinterwordspacing

\bibitem{leitner2019}
\BIBentryALTinterwordspacing
C.~Leitner, P.~A. Hager, H.~Penasso, M.~Tilp, L.~Benini, C.~Peham, and
  C.~Baumgartner, ``\BIBforeignlanguage{en}{Ultrasound as a {Tool} to {Study}
  {Muscle}–{Tendon} {Functions} during {Locomotion}: {A} {Systematic}
  {Review} of {Applications}},'' \emph{\BIBforeignlanguage{en}{Sensors}},
  vol.~19, no.~19, p. 4316, Jan. 2019, number: 19 Publisher: Multidisciplinary
  Digital Publishing Institute. [Online]. Available:
  \url{https://www.mdpi.com/1424-8220/19/19/4316}
\BIBentrySTDinterwordspacing

\bibitem{kruse_muscle_2018}
\BIBentryALTinterwordspacing
A.~Kruse, C.~Schranz, M.~Tilp, and M.~Svehlik, ``\BIBforeignlanguage{en}{Muscle
  and tendon morphology alterations in children and adolescents with mild forms
  of spastic cerebral palsy},'' \emph{\BIBforeignlanguage{en}{BMC Pediatrics}},
  vol.~18, no.~1, p. 156, May 2018. [Online]. Available:
  \url{https://doi.org/10.1186/s12887-018-1129-4}
\BIBentrySTDinterwordspacing

\bibitem{kenny2021}
\BIBentryALTinterwordspacing
J.-E.~S. Kenny, C.~E. Munding, J.~K. Eibl, A.~M. Eibl, B.~F. Long, A.~Boyes,
  J.~Yin, P.~Verrecchia, M.~Parrotta, R.~Gatzke, P.~A. Magnin, P.~N. Burns,
  F.~S. Foster, and C.~E.~M. Demore, ``\BIBforeignlanguage{en}{A novel,
  hands-free ultrasound patch for continuous monitoring of quantitative
  {Doppler} in the carotid artery},'' \emph{\BIBforeignlanguage{en}{Scientific
  Reports}}, vol.~11, no.~1, p. 7780, Apr. 2021, publisher: Nature Publishing
  Group. [Online]. Available:
  \url{https://www.nature.com/articles/s41598-021-87116-y}
\BIBentrySTDinterwordspacing

\bibitem{yang2024}
\BIBentryALTinterwordspacing
X.~Yang, C.~Castellini, D.~Farina, and H.~Liu, ``Ultrasound as a {Neurorobotic}
  {Interface}: {A} {Review},'' \emph{IEEE Transactions on Systems, Man, and
  Cybernetics: Systems}, vol.~54, no.~6, pp. 3534--3546, Jun. 2024, conference
  Name: IEEE Transactions on Systems, Man, and Cybernetics: Systems. [Online].
  Available: \url{https://ieeexplore.ieee.org/abstract/document/10436655}
\BIBentrySTDinterwordspacing

\bibitem{wang_continuous_2021}
\BIBentryALTinterwordspacing
C.~Wang, B.~Qi, M.~Lin, Z.~Zhang, M.~Makihata, B.~Liu, S.~Zhou, Y.-h. Huang,
  H.~Hu, Y.~Gu, Y.~Chen, Y.~Lei, T.~Lee, S.~Chien, K.-I. Jang, E.~B. Kistler,
  and S.~Xu, ``\BIBforeignlanguage{en}{Continuous monitoring of deep-tissue
  haemodynamics with stretchable ultrasonic phased arrays},''
  \emph{\BIBforeignlanguage{en}{Nature Biomedical Engineering}}, vol.~5, no.~7,
  pp. 749--758, Jul. 2021. [Online]. Available:
  \url{https://www.nature.com/articles/s41551-021-00763-4}
\BIBentrySTDinterwordspacing

\bibitem{du_conformable_2023}
\BIBentryALTinterwordspacing
W.~Du, L.~Zhang, E.~Suh, D.~Lin, C.~Marcus, L.~Ozkan, A.~Ahuja, S.~Fernandez,
  I.~I. Shuvo, D.~Sadat, W.~Liu, F.~Li, A.~P. Chandrakasan, T.~Ozmen, and
  C.~Dagdeviren, ``\BIBforeignlanguage{en}{Conformable ultrasound breast patch
  for deep tissue scanning and imaging},''
  \emph{\BIBforeignlanguage{en}{Science Advances}}, vol.~9, no.~30, p.
  eadh5325, Jul. 2023. [Online]. Available:
  \url{https://www.science.org/doi/10.1126/sciadv.adh5325}
\BIBentrySTDinterwordspacing

\bibitem{keller2023}
\BIBentryALTinterwordspacing
K.~Keller, C.~Leitner, C.~Baumgartner, L.~Benini, and F.~Greco,
  ``\BIBforeignlanguage{en}{Fully {Printed} {Flexible} {Ultrasound}
  {Transducer} for {Medical} {Applications}},''
  \emph{\BIBforeignlanguage{en}{Advanced Materials Technologies}}, vol.~8,
  no.~18, p. 2300577, 2023, \_eprint:
  https://onlinelibrary.wiley.com/doi/pdf/10.1002/admt.202300577. [Online].
  Available:
  \url{https://onlinelibrary.wiley.com/doi/abs/10.1002/admt.202300577}
\BIBentrySTDinterwordspacing

\bibitem{frey_wulpus_2022}
\BIBentryALTinterwordspacing
S.~Frey, S.~Vostrikov, L.~Benini, and A.~Cossettini, ``{WULPUS}: a {Wearable}
  {Ultra} {Low}-{Power} {Ultrasound} probe for multi-day monitoring of carotid
  artery and muscle activity,'' in \emph{2022 {IEEE} {International}
  {Ultrasonics} {Symposium} ({IUS})}.\hskip 1em plus 0.5em minus 0.4em\relax
  Venice, Italy: IEEE, Oct. 2022, pp. 1--4. [Online]. Available:
  \url{https://ieeexplore.ieee.org/document/9958156/}
\BIBentrySTDinterwordspacing

\bibitem{lin2024}
\BIBentryALTinterwordspacing
M.~Lin, Z.~Zhang, X.~Gao, Y.~Bian, R.~S. Wu, G.~Park, Z.~Lou, Z.~Zhang, X.~Xu,
  X.~Chen, A.~Kang, X.~Yang, W.~Yue, L.~Yin, C.~Wang, B.~Qi, S.~Zhou, H.~Hu,
  H.~Huang, M.~Li, Y.~Gu, J.~Mu, A.~Yang, A.~Yaghi, Y.~Chen, Y.~Lei, C.~Lu,
  R.~Wang, J.~Wang, S.~Xiang, E.~B. Kistler, N.~Vasconcelos, and S.~Xu,
  ``\BIBforeignlanguage{en}{A fully integrated wearable ultrasound system to
  monitor deep tissues in moving subjects},''
  \emph{\BIBforeignlanguage{en}{Nature Biotechnology}}, vol.~42, no.~3, pp.
  448--457, Mar. 2024, publisher: Nature Publishing Group. [Online]. Available:
  \url{https://www.nature.com/articles/s41587-023-01800-0}
\BIBentrySTDinterwordspacing

\bibitem{wang2019}
\BIBentryALTinterwordspacing
T.~Wang, D.~Zhang, L.~Wang, Y.~Zheng, T.~Gu, B.~Dorizzi, and X.~Zhou,
  ``Contactless {Respiration} {Monitoring} {Using} {Ultrasound} {Signal} {With}
  {Off}-the-{Shelf} {Audio} {Devices},'' \emph{IEEE Internet of Things
  Journal}, vol.~6, no.~2, pp. 2959--2973, Apr. 2019, conference Name: IEEE
  Internet of Things Journal. [Online]. Available:
  \url{https://ieeexplore.ieee.org/document/8502803}
\BIBentrySTDinterwordspacing

\bibitem{liu2023}
\BIBentryALTinterwordspacing
W.~Liu, S.~Chang, F.~Li, Y.~Xu, S.~Yan, and Y.~Liu, ``Contactless {Breathing}
  {Airflow} {Detection} on {Smartphone},'' \emph{IEEE Internet of Things
  Journal}, vol.~10, no.~4, pp. 3428--3439, Feb. 2023, conference Name: IEEE
  Internet of Things Journal. [Online]. Available:
  \url{https://ieeexplore.ieee.org/document/9951134}
\BIBentrySTDinterwordspacing

\bibitem{9219124}
Y.~Shen, H.~Zhang, Y.~Fan, A.~P. Lee, and L.~Xu, ``Smart health of ultrasound
  telemedicine based on deeply represented semantic segmentation,'' \emph{IEEE
  Internet of Things Journal}, vol.~8, no.~23, pp. 16\,770--16\,778, 2021.

\bibitem{mcintosh2017}
\BIBentryALTinterwordspacing
J.~McIntosh, A.~Marzo, M.~Fraser, and C.~Phillips, ``{EchoFlex}: {Hand}
  {Gesture} {Recognition} using {Ultrasound} {Imaging},'' in \emph{Proceedings
  of the 2017 {CHI} {Conference} on {Human} {Factors} in {Computing}
  {Systems}}, ser. {CHI} '17.\hskip 1em plus 0.5em minus 0.4em\relax New York,
  NY, USA: Association for Computing Machinery, May 2017, pp. 1923--1934.
  [Online]. Available: \url{https://dl.acm.org/doi/10.1145/3025453.3025807}
\BIBentrySTDinterwordspacing

\bibitem{vostrikov_complete_2023}
\BIBentryALTinterwordspacing
S.~Vostrikov, L.~Benini, and A.~Cossettini, ``Complete {Cardiorespiratory}
  {Monitoring} via {Wearable} {Ultra} {Low} {Power} {Ultrasound},'' in
  \emph{2023 {IEEE} {International} {Ultrasonics} {Symposium} ({IUS})}.\hskip
  1em plus 0.5em minus 0.4em\relax Montreal, QC, Canada: IEEE, Sep. 2023, pp.
  1--4. [Online]. Available:
  \url{https://ieeexplore.ieee.org/document/10307398/}
\BIBentrySTDinterwordspacing

\bibitem{AppleWat86:online}
``Apple watch series 9 - apple,''
  \url{https://www.apple.com/apple-watch-series-9/}, (Accessed on 07/26/2024).

\bibitem{frey_wearable_2023}
\BIBentryALTinterwordspacing
S.~Frey, V.~Kartsch, C.~Leitner, A.~Cossettini, S.~Vostrikov, S.~Benatti, and
  L.~Benini, ``A {Wearable} {Ultra}-{Low}-{Power} {sEMG}-triggered {Ultrasound}
  {System} for {Long}-term {Muscle} {Activity} {Monitoring},'' in \emph{2023
  {IEEE} {International} {Ultrasonics} {Symposium} ({IUS})}, Sep. 2023, pp.
  1--4, iSSN: 1948-5727. [Online]. Available:
  \url{https://ieeexplore.ieee.org/document/10307824}
\BIBentrySTDinterwordspacing

\bibitem{goldberger_front_2018}
\BIBentryALTinterwordspacing
A.~L. Goldberger, Z.~D. Goldberger, and A.~Shvilkin, Eds., \emph{Goldberger's
  {Clinical} {Electrocardiography}}, 9th~ed.\hskip 1em plus 0.5em minus
  0.4em\relax Boston: Elsevier, Jan. 2018. [Online]. Available:
  \url{https://www.sciencedirect.com/science/article/pii/B9780323401692000305}
\BIBentrySTDinterwordspacing

\bibitem{apple_take_nodate}
\BIBentryALTinterwordspacing
Apple, ``\BIBforeignlanguage{en}{Take an {ECG} with the {ECG} app on {Apple}
  {Watch}}.'' [Online]. Available: \url{https://support.apple.com/en-us/120278}
\BIBentrySTDinterwordspacing

\bibitem{spaccarotella_measurement_2021}
\BIBentryALTinterwordspacing
C.~A.~M. Spaccarotella, S.~Migliarino, A.~Mongiardo, J.~Sabatino, G.~Santarpia,
  S.~De~Rosa, A.~Curcio, and C.~Indolfi, ``\BIBforeignlanguage{en}{Measurement
  of the {QT} interval using the {Apple} {Watch}},''
  \emph{\BIBforeignlanguage{en}{Scientific Reports}}, vol.~11, no.~1, p. 10817,
  May 2021, publisher: Nature Publishing Group. [Online]. Available:
  \url{https://www.nature.com/articles/s41598-021-89199-z}
\BIBentrySTDinterwordspacing

\bibitem{fitbit_fitbit_nodate}
\BIBentryALTinterwordspacing
Fitbit, ``\BIBforeignlanguage{en}{Fitbit {ECG} app {\textbar} {Heart} {Rhythm}
  {Assessment}},'' (Accessed \today). [Online]. Available:
  \url{https://www.fitbit.com/global/dk/technology/ecg}
\BIBentrySTDinterwordspacing

\bibitem{samsung_measure_nodate}
\BIBentryALTinterwordspacing
Samsung, ``\BIBforeignlanguage{en-419}{Measure your {ECG} with the {Galaxy}
  {Watch} series},'' (Accessed \today). [Online]. Available:
  \url{https://www.samsung.com/latin_en/support/mobile-devices/measure-your-ecg-with-the-galaxy-watch-series/}
\BIBentrySTDinterwordspacing

\bibitem{garmin_ecg_nodate}
\BIBentryALTinterwordspacing
Garmin, ``\BIBforeignlanguage{en-US}{{ECG} {App} {\textbar} {Health} {Science}
  {\textbar} {Garmin} {Technology} {\textbar} {Garmin}},'' (Accessed \today).
  [Online]. Available:
  \url{https://www.garmin.com/en-US/garmin-technology/health-science/ecg/}
\BIBentrySTDinterwordspacing

\bibitem{polar_polar_nodate}
\BIBentryALTinterwordspacing
Polar, ``\BIBforeignlanguage{en}{Polar {Technologies} {\textbar} {Wrist}-{ECG}
  {\textbar} {Polar} {Global}},'' (Accessed \today). [Online]. Available:
  \url{https://www.polar.com/en/explore/heart/ecg/}
\BIBentrySTDinterwordspacing

\bibitem{allen_photoplethysmography_2022}
\BIBentryALTinterwordspacing
J.~Allen and P.~Kyriacou, Eds., \emph{Photoplethysmography {Technology},
  {Signal} {Analysis} and {Applications}}.\hskip 1em plus 0.5em minus
  0.4em\relax Academic Press, Jan. 2022. [Online]. Available:
  \url{https://www.sciencedirect.com/science/article/pii/B9780128233740000918}
\BIBentrySTDinterwordspacing

\bibitem{de_pinho_ferreira_review_2021}
\BIBentryALTinterwordspacing
N.~De~Pinho~Ferreira, C.~Gehin, and B.~Massot, ``A {Review} of {Methods} for
  {Non}-{Invasive} {Heart} {Rate} {Measurement} on {Wrist},'' \emph{IRBM},
  vol.~42, no.~1, pp. 4--18, Feb. 2021. [Online]. Available:
  \url{https://www.sciencedirect.com/science/article/pii/S1959031820300804}
\BIBentrySTDinterwordspacing

\bibitem{ebrahimi_ultralow-power_2023}
\BIBentryALTinterwordspacing
Z.~Ebrahimi and B.~Gosselin, ``Ultralow-{Power} {Photoplethysmography} ({PPG})
  {Sensors}: {A} {Methodological} {Review},'' \emph{IEEE Sensors Journal},
  vol.~23, no.~15, pp. 16\,467--16\,480, Aug. 2023, conference Name: IEEE
  Sensors Journal. [Online]. Available:
  \url{https://ieeexplore.ieee.org/abstract/document/10153489}
\BIBentrySTDinterwordspacing

\bibitem{apple_monitor_nodate}
\BIBentryALTinterwordspacing
Apple, ``\BIBforeignlanguage{en}{Monitor your heart rate with {Apple}
  {Watch}},'' (Accessed \today). [Online]. Available:
  \url{https://support.apple.com/en-us/120277}
\BIBentrySTDinterwordspacing

\bibitem{analog_devices_max86150_nodate}
\BIBentryALTinterwordspacing
A.~Devices, ``{MAX86150} {Datasheet} and {Product} {Info} {\textbar} {Analog}
  {Devices},'' (Accessed \today). [Online]. Available:
  \url{https://www.analog.com/en/products/max86150.html}
\BIBentrySTDinterwordspacing

\bibitem{texas_instuments_afe4900_nodate}
T.~Instuments, ``{AFE4900} data sheet, product information and support
  {\textbar} {TI}.com,'' (Accessed \today).

\bibitem{ray_review_2023}
\BIBentryALTinterwordspacing
D.~Ray, T.~Collins, S.~I. Woolley, and P.~V.~S. Ponnapalli, ``A {Review} of
  {Wearable} {Multi}-{Wavelength} {Photoplethysmography},'' \emph{IEEE Reviews
  in Biomedical Engineering}, vol.~16, pp. 136--151, 2023, conference Name:
  IEEE Reviews in Biomedical Engineering. [Online]. Available:
  \url{https://ieeexplore.ieee.org/document/9582790}
\BIBentrySTDinterwordspacing

\bibitem{biswas_heart_2019}
\BIBentryALTinterwordspacing
D.~Biswas, N.~Simões-Capela, C.~Van~Hoof, and N.~Van~Helleputte, ``Heart
  {Rate} {Estimation} {From} {Wrist}-{Worn} {Photoplethysmography}: {A}
  {Review},'' \emph{IEEE Sensors Journal}, vol.~19, no.~16, pp. 6560--6570,
  Aug. 2019, conference Name: IEEE Sensors Journal. [Online]. Available:
  \url{https://ieeexplore.ieee.org/document/8703846}
\BIBentrySTDinterwordspacing

\bibitem{loh_application_2022}
\BIBentryALTinterwordspacing
H.~W. Loh, S.~Xu, O.~Faust, C.~P. Ooi, P.~D. Barua, S.~Chakraborty, R.-S. Tan,
  F.~Molinari, and U.~R. Acharya, ``Application of photoplethysmography signals
  for healthcare systems: {An} in-depth review,'' \emph{Computer Methods and
  Programs in Biomedicine}, vol. 216, p. 106677, Apr. 2022. [Online].
  Available:
  \url{https://www.sciencedirect.com/science/article/pii/S0169260722000621}
\BIBentrySTDinterwordspacing

\bibitem{koerber2023accuracy}
\BIBentryALTinterwordspacing
D.~Koerber, S.~Khan, T.~Shamsheri, A.~Kirubarajan, and S.~Mehta, ``Accuracy of
  heart rate measurement with wrist-worn wearable devices in various skin
  tones: a systematic review,'' \emph{Journal of Racial and Ethnic Health
  Disparities}, vol.~10, no.~6, pp. 2676--2684, 2023. [Online]. Available:
  \url{https://link.springer.com/article/10.1007/s40615-022-01446-9}
\BIBentrySTDinterwordspacing

\bibitem{nelson2020guidelines}
\BIBentryALTinterwordspacing
B.~W. Nelson, C.~A. Low, N.~Jacobson, P.~Are{\'a}n, J.~Torous, and N.~B. Allen,
  ``Guidelines for wrist-worn consumer wearable assessment of heart rate in
  biobehavioral research,'' \emph{NPJ digital medicine}, vol.~3, no.~1, p.~90,
  2020. [Online]. Available:
  \url{https://www.nature.com/articles/s41746-020-0297-4}
\BIBentrySTDinterwordspacing

\bibitem{long_wearable_2022}
\BIBentryALTinterwordspacing
N.~M.~H. Long and W.-Y. Chung, ``Wearable {Wrist} {Photoplethysmography} for
  {Optimal} {Monitoring} of {Vital} {Signs}: {A} {Unified} {Perspective} on
  {Pulse} {Waveforms},'' \emph{IEEE Photonics Journal}, vol.~14, no.~2, pp.
  1--17, Apr. 2022, conference Name: IEEE Photonics Journal. [Online].
  Available: \url{https://ieeexplore.ieee.org/document/9720222}
\BIBentrySTDinterwordspacing

\bibitem{hahnen_accuracy_2020}
\BIBentryALTinterwordspacing
C.~Hahnen, C.~G. Freeman, N.~Haldar, J.~N. Hamati, D.~M. Bard, V.~Murali, G.~J.
  Merli, J.~I. Joseph, and N.~v. Helmond, ``\BIBforeignlanguage{EN}{Accuracy of
  {Vital} {Signs} {Measurements} by a {Smartwatch} and a {Portable} {Health}
  {Device}: {Validation} {Study}},'' \emph{\BIBforeignlanguage{EN}{JMIR mHealth
  and uHealth}}, vol.~8, no.~2, p. e16811, Feb. 2020, company: JMIR mHealth and
  uHealth Distributor: JMIR mHealth and uHealth Institution: JMIR mHealth and
  uHealth Label: JMIR mHealth and uHealth Publisher: JMIR Publications Inc.,
  Toronto, Canada. [Online]. Available:
  \url{https://mhealth.jmir.org/2020/2/e16811}
\BIBentrySTDinterwordspacing

\bibitem{sarhaddi_comprehensive_2022}
\BIBentryALTinterwordspacing
F.~Sarhaddi, K.~Kazemi, I.~Azimi, R.~Cao, H.~Niela-Vilén, A.~Axelin,
  P.~Liljeberg, and A.~M. Rahmani, ``\BIBforeignlanguage{en}{A comprehensive
  accuracy assessment of {Samsung} smartwatch heart rate and heart rate
  variability},'' \emph{\BIBforeignlanguage{en}{PLOS ONE}}, vol.~17, no.~12, p.
  e0268361, Dec. 2022, publisher: Public Library of Science. [Online].
  Available:
  \url{https://journals.plos.org/plosone/article?id=10.1371/journal.pone.0268361}
\BIBentrySTDinterwordspacing

\bibitem{hajj-boutros_wrist-worn_2023}
\BIBentryALTinterwordspacing
G.~Hajj-Boutros, M.-A. Landry-Duval, A.~S. Comtois, G.~Gouspillou, and A.~D.
  Karelis, ``Wrist-worn devices for the measurement of heart rate and energy
  expenditure: {A} validation study for the {Apple} {Watch} 6, {Polar}
  {Vantage} {V} and {Fitbit} {Sense},'' \emph{European Journal of Sport
  Science}, vol.~23, no.~2, pp. 165--177, Feb. 2023, publisher: Routledge.
  [Online]. Available: \url{https://doi.org/10.1080/17461391.2021.2023656}
\BIBentrySTDinterwordspacing

\bibitem{fang_wrist_2021}
\BIBentryALTinterwordspacing
P.~Fang, Y.~Peng, W.-H. Lin, Y.~Wang, S.~Wang, X.~Zhang, K.~Wu, and G.~Li,
  ``Wrist {Pulse} {Recording} {With} a {Wearable}
  {Piezoresistor}-{Piezoelectret} {Compound} {Sensing} {System} and {Its}
  {Applications} in {Health} {Monitoring},'' \emph{IEEE Sensors Journal},
  vol.~21, no.~18, pp. 20\,921--20\,930, Sep. 2021, conference Name: IEEE
  Sensors Journal. [Online]. Available:
  \url{https://ieeexplore.ieee.org/document/9475051}
\BIBentrySTDinterwordspacing

\bibitem{polley_wearable_2021}
\BIBentryALTinterwordspacing
C.~Polley, T.~Jayarathna, U.~Gunawardana, G.~Naik, T.~Hamilton, E.~Andreozzi,
  P.~Bifulco, D.~Esposito, J.~Centracchio, and G.~Gargiulo,
  ``\BIBforeignlanguage{en}{Wearable {Bluetooth} {Triage} {Healthcare}
  {Monitoring} {System}},'' \emph{\BIBforeignlanguage{en}{Sensors}}, vol.~21,
  no.~22, p. 7586, Jan. 2021, number: 22 Publisher: Multidisciplinary Digital
  Publishing Institute. [Online]. Available:
  \url{https://www.mdpi.com/1424-8220/21/22/7586}
\BIBentrySTDinterwordspacing

\bibitem{pui_audio_benders_nodate}
\BIBentryALTinterwordspacing
P.~Audio, ``\BIBforeignlanguage{en-US}{Benders {\textbar}
  {AB1290B}-{LW100}-{R}},'' (Accessed \today). [Online]. Available:
  \url{https://puiaudio.com/product/benders/ab1290b-lw100-r}
\BIBentrySTDinterwordspacing

\bibitem{barnes_pinkysil_nodate}
\BIBentryALTinterwordspacing
Barnes, ``\BIBforeignlanguage{en-US}{Pinkysil {Fast} {Setting} {Silicone}},''
  (Accessed \today). [Online]. Available:
  \url{https://www.barnes.com.au/single-product/pinkysil-fast-set-silicone/}
\BIBentrySTDinterwordspacing

\bibitem{peng_noninvasive_2021}
\BIBentryALTinterwordspacing
C.~Peng, M.~Chen, H.~K. Sim, Y.~Zhu, and X.~Jiang, ``Noninvasive and
  {Nonocclusive} {Blood} {Pressure} {Monitoring} via a {Flexible}
  {Piezo}-{Composite} {Ultrasonic} {Sensor},'' \emph{IEEE Sensors Journal},
  vol.~21, no.~3, pp. 2642--2650, Feb. 2021, conference Name: IEEE Sensors
  Journal. [Online]. Available:
  \url{https://ieeexplore.ieee.org/document/9186675}
\BIBentrySTDinterwordspacing

\bibitem{park_opto-ultrasound_2022}
\BIBentryALTinterwordspacing
J.~Park, B.~Park, J.~Ahn, D.~Kim, J.~Y. Kim, H.~H. Kim, and C.~Kim,
  ``\BIBforeignlanguage{EN}{Opto-ultrasound biosensor for wearable and mobile
  devices: realization with a transparent ultrasound transducer},''
  \emph{\BIBforeignlanguage{EN}{Biomedical Optics Express}}, vol.~13, no.~9,
  pp. 4684--4692, Sep. 2022, publisher: Optica Publishing Group. [Online].
  Available: \url{https://opg.optica.org/boe/abstract.cfm?uri=boe-13-9-4684}
\BIBentrySTDinterwordspacing

\bibitem{shumba_monitoring_2024}
\BIBentryALTinterwordspacing
A.~T. Shumba, S.~M. Demir, V.~M. Mastronardi, F.~Rizzi, G.~De~Marzo,
  L.~Fachechi, P.~M. Ros, D.~Demarchi, L.~Patrono, and M.~De~Vittorio,
  ``Monitoring {Cardiovascular} {Physiology} {Using} {Bio}-{Compatible} {AlN}
  {Piezoelectric} {Skin} {Sensors},'' \emph{IEEE Access}, vol.~12, pp.
  16\,951--16\,962, 2024, conference Name: IEEE Access. [Online]. Available:
  \url{https://ieeexplore.ieee.org/document/10414985}
\BIBentrySTDinterwordspacing

\bibitem{formlabs_elastic_nodate}
\BIBentryALTinterwordspacing
formlabs, ``\BIBforeignlanguage{en}{Elastic {50A} {Resin} {V2}},'' (Accessed
  \today). [Online]. Available:
  \url{https://formlabs.com/ch/shop/materials/elastic-50a-resin-v2/}
\BIBentrySTDinterwordspacing

\bibitem{piezo_hannas_customize_nodate}
\BIBentryALTinterwordspacing
P.~Hannas, ``\BIBforeignlanguage{en}{Customize {10Mhz} a-scan ultrasound
  probes},'' (Accessed \today). [Online]. Available:
  \url{https://www.piezohannas.com}
\BIBentrySTDinterwordspacing

\bibitem{henriksen_using_2018}
\BIBentryALTinterwordspacing
A.~Henriksen, M.~H. Mikalsen, A.~Z. Woldaregay, M.~Muzny, G.~Hartvigsen, L.~A.
  Hopstock, and S.~Grimsgaard, ``\BIBforeignlanguage{EN}{Using {Fitness}
  {Trackers} and {Smartwatches} to {Measure} {Physical} {Activity} in
  {Research}: {Analysis} of {Consumer} {Wrist}-{Worn} {Wearables}},''
  \emph{\BIBforeignlanguage{EN}{Journal of Medical Internet Research}},
  vol.~20, no.~3, p. e9157, Mar. 2018, company: Journal of Medical Internet
  Research Distributor: Journal of Medical Internet Research Institution:
  Journal of Medical Internet Research Label: Journal of Medical Internet
  Research Publisher: JMIR Publications Inc., Toronto, Canada. [Online].
  Available: \url{https://www.jmir.org/2018/3/e110}
\BIBentrySTDinterwordspacing

\bibitem{islam_visualizing_2020}
\BIBentryALTinterwordspacing
A.~Islam, A.~Bezerianos, B.~Lee, T.~Blascheck, and P.~Isenberg, ``Visualizing
  {Information} on {Watch} {Faces}: {A} {Survey} with {Smartwatch} {Users},''
  in \emph{2020 {IEEE} {Visualization} {Conference} ({VIS})}, Oct. 2020, pp.
  156--160. [Online]. Available:
  \url{https://ieeexplore.ieee.org/abstract/document/9331267}
\BIBentrySTDinterwordspacing

\bibitem{nelson_guidelines_2020}
\BIBentryALTinterwordspacing
B.~W. Nelson, C.~A. Low, N.~Jacobson, P.~Areán, J.~Torous, and N.~B. Allen,
  ``\BIBforeignlanguage{en}{Guidelines for wrist-worn consumer wearable
  assessment of heart rate in biobehavioral research},''
  \emph{\BIBforeignlanguage{en}{npj Digital Medicine}}, vol.~3, no.~1, pp.
  1--9, Jun. 2020, publisher: Nature Publishing Group. [Online]. Available:
  \url{https://www.nature.com/articles/s41746-020-0297-4}
\BIBentrySTDinterwordspacing

\bibitem{jiang_emerging_2022}
\BIBentryALTinterwordspacing
S.~Jiang, P.~Kang, X.~Song, B.~P. Lo, and P.~B. Shull, ``Emerging {Wearable}
  {Interfaces} and {Algorithms} for {Hand} {Gesture} {Recognition}: {A}
  {Survey},'' \emph{IEEE Reviews in Biomedical Engineering}, vol.~15, pp.
  85--102, 2022, conference Name: IEEE Reviews in Biomedical Engineering.
  [Online]. Available: \url{https://ieeexplore.ieee.org/document/9426433}
\BIBentrySTDinterwordspacing

\bibitem{suh_worker_2024}
\BIBentryALTinterwordspacing
S.~Suh, V.~F. Rey, S.~Bian, Y.-C. Huang, J.~M. Rožanec, H.~T. Ghinani,
  B.~Zhou, and P.~Lukowicz, ``Worker {Activity} {Recognition} in
  {Manufacturing} {Line} {Using} {Near}-{Body} {Electric} {Field},'' \emph{IEEE
  Internet of Things Journal}, vol.~11, no.~7, pp. 11\,554--11\,565, Apr. 2024,
  conference Name: IEEE Internet of Things Journal. [Online]. Available:
  \url{https://ieeexplore.ieee.org/document/10308956}
\BIBentrySTDinterwordspacing

\bibitem{swatch_replacement_nodate}
\BIBentryALTinterwordspacing
Swatch, ``\BIBforeignlanguage{en}{Replacement straps for your {Swatch} watches
  {\textbar} {Swatch}® {Switzerland}},'' (Accessed \today). [Online].
  Available: \url{https://www.swatch.com/en-ch/accessories/watch-straps/}
\BIBentrySTDinterwordspacing

\bibitem{benini_wireless_2006}
\BIBentryALTinterwordspacing
L.~Benini, E.~Farella, and C.~Guiducci, ``Wireless sensor networks: {Enabling}
  technology for ambient intelligence,'' \emph{Microelectronics Journal},
  vol.~37, no.~12, pp. 1639--1649, Dec. 2006. [Online]. Available:
  \url{https://www.sciencedirect.com/science/article/pii/S0026269206001728}
\BIBentrySTDinterwordspacing

\bibitem{vostrikov_towards_2021}
\BIBentryALTinterwordspacing
S.~Vostrikov, A.~Cossettini, C.~Vogt, C.~Leitner, M.~Magno, and L.~Benini,
  ``Towards an {Open}, {Flexible}, {Wearable} {Ultrasound} {Probe} for
  {Musculoskeletal} {Monitoring},'' in \emph{2021 {IEEE} {International}
  {Ultrasonics} {Symposium} ({IUS})}.\hskip 1em plus 0.5em minus 0.4em\relax
  Xi'an, China: IEEE, Sep. 2021, pp. 1--4. [Online]. Available:
  \url{https://ieeexplore.ieee.org/document/9593910/}
\BIBentrySTDinterwordspacing

\bibitem{yang2019wearable}
\BIBentryALTinterwordspacing
X.~Yang, Z.~Chen, N.~Hettiarachchi, J.~Yan, and H.~Liu, ``A wearable ultrasound
  system for sensing muscular morphological deformations,'' \emph{IEEE
  Transactions on Systems, Man, and Cybernetics: Systems}, vol.~51, no.~6, pp.
  3370--3379, 2019. [Online]. Available:
  \url{https://ieeexplore.ieee.org/document/8760415}
\BIBentrySTDinterwordspacing

\bibitem{song2019design}
\BIBentryALTinterwordspacing
I.~Song, J.~Yoon, J.~Kang, M.~Kim, W.~S. Jang, N.-Y. Shin, and Y.~Yoo, ``Design
  and implementation of a new wireless carotid neckband doppler system with
  wearable ultrasound sensors: Preliminary results,'' \emph{Applied Sciences},
  vol.~9, no.~11, p. 2202, 2019. [Online]. Available:
  \url{https://www.mdpi.com/2076-3417/9/11/2202}
\BIBentrySTDinterwordspacing

\bibitem{fournelle2021portable}
\BIBentryALTinterwordspacing
M.~Fournelle, T.~Gr{\"u}n, D.~Speicher, S.~Weber, M.~Yilmaz, D.~Schoeb,
  A.~Miernik, G.~Reis, S.~Tretbar, and H.~Hewener, ``Portable ultrasound
  research system for use in automated bladder monitoring with
  machine-learning-based segmentation,'' \emph{Sensors}, vol.~21, no.~19, p.
  6481, 2021. [Online]. Available:
  \url{https://www.mdpi.com/1424-8220/21/19/6481}
\BIBentrySTDinterwordspacing

\bibitem{yin2022wearable}
\BIBentryALTinterwordspacing
Z.~Yin, H.~Chen, X.~Yang, Y.~Liu, N.~Zhang, J.~Meng, and H.~Liu, ``A wearable
  ultrasound interface for prosthetic hand control,'' \emph{IEEE journal of
  biomedical and health informatics}, vol.~26, no.~11, pp. 5384--5393, 2022.
  [Online]. Available: \url{https://ieeexplore.ieee.org/document/9872106}
\BIBentrySTDinterwordspacing

\bibitem{st_microelectronics_stm32l4_nodate}
\BIBentryALTinterwordspacing
S.~Microelectronics, ``\BIBforeignlanguage{en}{{STM32L4} - {ARM} {Cortex}-{M4}
  ultra-low-power {MCUs} - {STMicroelectronics}},'' (Accessed \today).
  [Online]. Available:
  \url{https://www.st.com/en/microcontrollers-microprocessors/stm32l4-series.html}
\BIBentrySTDinterwordspacing

\bibitem{microchip_pic32mz1024efh064_nodate}
\BIBentryALTinterwordspacing
Microchip, ``\BIBforeignlanguage{en-US}{{PIC32MZ1024EFH064}},'' (Accessed
  \today). [Online]. Available:
  \url{https://www.microchip.com/en-us/product/pic32mz1024efh064}
\BIBentrySTDinterwordspacing

\bibitem{texas_instuments_msp430fr5043_nodate}
\BIBentryALTinterwordspacing
T.~Instuments, ``{MSP430FR5043} data sheet, product information and support
  {\textbar} {TI}.com,'' (Accessed \today). [Online]. Available:
  \url{https://www.ti.com/product/MSP430FR5043}
\BIBentrySTDinterwordspacing

\bibitem{boni_ultrasound_2018}
\BIBentryALTinterwordspacing
E.~Boni, A.~C.~H. Yu, S.~Freear, J.~A. Jensen, and P.~Tortoli, ``Ultrasound
  {Open} {Platforms} for {Next}-{Generation} {Imaging} {Technique}
  {Development},'' \emph{IEEE Transactions on Ultrasonics, Ferroelectrics, and
  Frequency Control}, vol.~65, no.~7, pp. 1078--1092, Jul. 2018, conference
  Name: IEEE Transactions on Ultrasonics, Ferroelectrics, and Frequency
  Control. [Online]. Available:
  \url{https://ieeexplore.ieee.org/document/8374071}
\BIBentrySTDinterwordspacing

\bibitem{kong_edge_2022}
\BIBentryALTinterwordspacing
X.~Kong, Y.~Wu, H.~Wang, and F.~Xia, ``Edge {Computing} for {Internet} of
  {Everything}: {A} {Survey},'' \emph{IEEE Internet of Things Journal}, vol.~9,
  no.~23, pp. 23\,472--23\,485, Dec. 2022, conference Name: IEEE Internet of
  Things Journal. [Online]. Available:
  \url{https://ieeexplore.ieee.org/document/9863881}
\BIBentrySTDinterwordspacing

\bibitem{giordano_towards_2023}
\BIBentryALTinterwordspacing
M.~Giordano, K.~Keller, F.~Greco, L.~Benini, M.~Magno, and C.~Leitner,
  ``Towards a {Novel} {Ultrasound} {System} {Based} on {Low}-{Frequency}
  {Feature} {Extraction} {From} a {Fully}-{Printed} {Flexible} {Transducer},''
  in \emph{2023 {IEEE} {Biomedical} {Circuits} and {Systems} {Conference}
  ({BioCAS})}.\hskip 1em plus 0.5em minus 0.4em\relax Toronto, ON, Canada:
  IEEE, Oct. 2023, pp. 1--5. [Online]. Available:
  \url{https://ieeexplore.ieee.org/document/10388792/}
\BIBentrySTDinterwordspacing

\bibitem{merenda_edge_2020}
\BIBentryALTinterwordspacing
M.~Merenda, C.~Porcaro, and D.~Iero, ``\BIBforeignlanguage{en}{Edge {Machine}
  {Learning} for {AI}-{Enabled} {IoT} {Devices}: {A} {Review}},''
  \emph{\BIBforeignlanguage{en}{Sensors}}, vol.~20, no.~9, p. 2533, Jan. 2020,
  number: 9 Publisher: Multidisciplinary Digital Publishing Institute.
  [Online]. Available: \url{https://www.mdpi.com/1424-8220/20/9/2533}
\BIBentrySTDinterwordspacing

\bibitem{waasdorp_combining_2021}
\BIBentryALTinterwordspacing
R.~Waasdorp, W.~Mugge, H.~J. Vos, J.~H. de~Groot, M.~D. Verweij, N.~de~Jong,
  A.~C. Schouten, and V.~Daeichin, ``Combining {Ultrafast} {Ultrasound} and
  {High}-{Density} {EMG} to {Assess} {Local} {Electromechanical} {Muscle}
  {Dynamics}: {A} {Feasibility} {Study},'' \emph{IEEE Access}, vol.~9, pp.
  45\,277--45\,288, 2021, conference Name: IEEE Access. [Online]. Available:
  \url{https://ieeexplore.ieee.org/document/9381253/?arnumber=9381253}
\BIBentrySTDinterwordspacing

\bibitem{canning_individuals_2014}
\BIBentryALTinterwordspacing
K.~L. Canning, R.~E. Brown, V.~K. Jamnik, A.~Salmon, C.~I. Ardern, and J.~L.
  Kuk, ``\BIBforeignlanguage{en}{Individuals {Underestimate} {Moderate} and
  {Vigorous} {Intensity} {Physical} {Activity}},''
  \emph{\BIBforeignlanguage{en}{PLOS ONE}}, vol.~9, no.~5, p. e97927, May 2014,
  publisher: Public Library of Science. [Online]. Available:
  \url{https://journals.plos.org/plosone/article?id=10.1371/journal.pone.0097927}
\BIBentrySTDinterwordspacing

\bibitem{la_flexible_2022}
\BIBentryALTinterwordspacing
T.-G. La and L.~H. Le, ``\BIBforeignlanguage{en}{Flexible and {Wearable}
  {Ultrasound} {Device} for {Medical} {Applications}: {A} {Review} on
  {Materials}, {Structural} {Designs}, and {Current} {Challenges}},''
  \emph{\BIBforeignlanguage{en}{Advanced Materials Technologies}}, vol.~7,
  no.~3, p. 2100798, 2022, \_eprint:
  https://onlinelibrary.wiley.com/doi/pdf/10.1002/admt.202100798. [Online].
  Available:
  \url{https://onlinelibrary.wiley.com/doi/abs/10.1002/admt.202100798}
\BIBentrySTDinterwordspacing

\bibitem{nordicsemiPowerProfiler}
``{P}ower {P}rofiler {K}it {I}{I} --- nordicsemi.com,''
  \url{https://www.nordicsemi.com/Products/Development-hardware/Power-Profiler-Kit-2},
  [Accessed 05-06-2024].

\bibitem{comaprison_ppg}
\BIBentryALTinterwordspacing
B.~W. Nelson and N.~B. Allen, ``Accuracy of consumer wearable heart rate
  measurement during an ecologically valid 24-hour period: Intraindividual
  validation study,'' \emph{JMIR Mhealth Uhealth}, vol.~7, no.~3, p. e10828,
  Mar 2019. [Online]. Available: \url{https://mhealth.jmir.org/2019/3/e10828/}
\BIBentrySTDinterwordspacing

\bibitem{kral2023slower}
\BIBentryALTinterwordspacing
T.~R. Kral, H.~Y. Weng, V.~Mitra, T.~P. Imhoff-Smith, E.~Azemi, R.~I. Goldman,
  M.~A. Rosenkranz, S.~Wu, A.~Chen, and R.~J. Davidson, ``Slower respiration
  rate is associated with higher self-reported well-being after wellness
  training,'' \emph{Scientific Reports}, vol.~13, no.~1, p. 15953, 2023.
  [Online]. Available: \url{https://www.nature.com/articles/s41598-023-43176-w}
\BIBentrySTDinterwordspacing

\bibitem{AppleWat87:online}
``Apple watch series 7 battery replacement - ifixit repair guide,''
  \url{https://www.ifixit.com/Guide/Apple+Watch+Series+7+Battery+Replacement/152655},
  (Accessed on 07/11/2024).

\bibitem{8478346}
\BIBentryALTinterwordspacing
A.~M. Zakrzewski, A.~Y. Huang, R.~Zubajlo, and B.~W. Anthony, ``Real-time blood
  pressure estimation from force-measured ultrasound,'' \emph{IEEE Transactions
  on Biomedical Engineering}, vol.~65, no.~11, pp. 2405--2416, 2018. [Online].
  Available: \url{https://ieeexplore.ieee.org/document/8478346}
\BIBentrySTDinterwordspacing

\bibitem{park2020ultrasound}
\BIBentryALTinterwordspacing
E.-Y. Park, J.~Baik, H.~Kim, S.-M. Park, and C.~Kim, ``Ultrasound-modulated
  optical glucose sensing using a 1645 nm laser,'' \emph{Scientific Reports},
  vol.~10, no.~1, p. 13361, 2020. [Online]. Available:
  \url{https://www.nature.com/articles/s41598-020-70305-6}
\BIBentrySTDinterwordspacing

\end{thebibliography}

\end{document}